\documentclass[twocolumn]{aa}

\usepackage[varg]{txfonts}
\include{bookdefs}
\usepackage{graphicx}
\usepackage{xspace}
\usepackage{amsmath}
\usepackage{placeins}
\usepackage{gensymb}
\usepackage{color}
\usepackage{float}
\usepackage[document]{ragged2e}
\usepackage{todonotes}

\renewcommand{\vec}[1]{\mbox{\boldmath$#1$}}
\newcommand{\music}{\texttt{MUSIC}\xspace}

\definecolor{orange}{rgb}{.9,.3,0}

\begin{document}
\title{Comparison of two- and three-dimensional compressible convection in a pre-main sequence star}
\author{J. Pratt\inst{\ref{inst2},\ref{inst1}}  \and I. Baraffe\inst{\ref{inst2},\ref{inst3}} \and  T. Goffrey\inst{\ref{inst2},\ref{inst4}} \and  C. Geroux\inst{\ref{inst5}}  \and  T. Constantino\inst{\ref{inst2}} \and D. Folini\inst{\ref{inst3}}  \and R. Walder\inst{\ref{inst3}}  }
\institute{Astrophysics, College of Engineering, Mathematics and Physical Sciences, University of Exeter, EX4 4QL Exeter, United Kingdom \label{inst2} 
\and Department of Physics and Astronomy, Georgia State University, Atlanta GA 30303, USA \label{inst1} 
\and \'Ecole Normale Sup\'erieure de Lyon, CRAL (UMR CNRS 5574), Universit\'e de Lyon 1, 69007 Lyon, France\label{inst3} 
\and Centre for Fusion, Space and Astrophysics, Department of Physics, University of Warwick, Coventy, CV4 7AL, United Kingdom \label{inst4}
\and ACENET/ Dalhousie University, Halifax, Nova Scotia, Canada\label{inst5} } 

\titlerunning{Dimensional comparison}
\authorrunning{J. Pratt et. al.}

\abstract
 {A one-dimensional description of stellar dynamics is at the basis of stellar evolution modeling.  Designed to investigate open problems in stellar evolution, the MUltidimensional Stellar Implicit Code (\music) expands a realistic one-dimensional profile of a star's internal structure to examine the interior dynamics of a specific star through either two- or three-dimensional hydrodynamic simulations.}
{Extending our recent studies of two-dimensional stellar convection to 3D, we compare three-dimensional hydrodynamic simulations to identically set-up two-dimensional simulations, for the realistic pre-main sequence star that we examined in \citet{prattspherical,pratt2017extreme}.}
{We compare statistical quantities related to convective flows including: average velocity, vorticity, local enstrophy, and penetration depth beneath a convection zone.  These statistics are produced during stationary, steady-state compressible convection in the star's convection zone.}
{Our simulations confirm the common result that two-dimensional simulations of stellar convection have a higher magnitude of velocity on average than three-dimensional simulations.  As we found in \citet{prattspherical}, boundary conditions and the extent of the spherical shell can affect the magnitude and variability of convective velocities.  The difference between 2D and 3D velocities is dependent on these background points; in our simulations this can have an effect as large as the difference resulting from the dimensionality of the simulation.  Nevertheless, radial velocities near the convective boundary are comparable in our 2D and 3D simulations.  
The average local enstrophy of the flow is lower for two-dimensional simulations than for three-dimensional simulations, indicating a different shape and structuring of 3D stellar convection.  
We perform a statistical analysis of the depth of convective penetration below the convection zone, using the model proposed in our recent study \citep{pratt2017extreme}.
That statistical model was developed based on two-dimensional simulations, which allowed us to examine longer times and higher radial resolution than are possible in 3D.  Here we analyze the convective penetration in three dimensional simulations, and compare the results to identically set-up 2D simulations.  In 3D the penetration depth is as large as the penetration depth calculated from 2D simulations. }
{}
\maketitle

\section{Introduction}

Studies of stellar convection have long reported that two-dimensional fluid simulations result in higher velocities than three-dimensional simulations \citep{muthsam1995numerical,meakin2007turbulent,arnett2011toward}.  Still two-dimensional modeling of convection remains a useful tool in the 321D link, i.e. the effort to improve one-dimensional stellar evolution modeling using two- and three-dimensional stellar hydrodynamics simulations.  Two-dimensional simulations allow the simultaneous exploration of longer times and higher radial resolutions\footnote{Higher radial resolution allows the stratification and temperature gradients of the star to be reproduced in a way that is more accurate to the 1D stellar structure.} than three-dimensional simulations.   The MUltidimensional Stellar Implicit Code (\music) is a stellar hydrodynamics code that has been designed to work in concert with one-dimensional stellar evolution calculations.  \music has been extensively tested for two-dimensional stellar convection in spherical shells of different radial extent and for different boundary conditions \citep{prattspherical}, benchmarked against other codes for fundamental hydrodynamic test problems \citep{goffrey2017benchmarking}, and tested for accuracy for low Mach number flows \citep{maximepaper}.  The aim of the present work is to quantify differences between two-dimensional and three-dimensional stellar convection, in order to extend our recent studies \citep{prattspherical, pratt2017extreme} to 3D.  This work provides a detailed comparison based on robust statistics, helpful for designing and interpreting future studies that use the \music code or other implicit large-eddy simulations of global stellar convection.  A detailed comparison of dimensionality is also useful in establishing the 321D link.

Two-dimensional turbulent flows are known to be fundamentally different from three-dimensional turbulent flows because an 
inverse energy cascade operates in 2D turbulence \citep[e.g.][]{kraichnan1971inertial,batchelor1969computation}.   This inverse cascade means that small-scale hydrodynamic motions can feed back on the largest scales of the flow.
As a consequence, studies of turbulent convection, including ``box-in-a-star'' simulations of fingering convection \citep[e.g.][]{garaud20152d}, have found that substantial differences can arise between simulations performed in two- and three-dimensions.  However for simulations on the scale of a stellar convection zone, both 2D and 3D simulations are typically dominated by large-scale coherent structures like plumes and convection rolls \citep[e.g. see visualizations in][]{alvan2015characterizing,pratt2017extreme,kapyla2018effects,strugarek2016modeling}.  The smaller scales relevant to turbulence may be damped by a dissipation that is significantly larger than is realistic in stars, modeled as in a large-eddy simulation (LES), or simply not included in the simulation because they are smaller than the grid scale as in an implicit large-eddy simulation (ILES) \citep[as discussed by][]{garnier2009large,grinstein2007implicit}. 
The differences that exist between 2D and 3D turbulent flows have an unknown impact on the large-scale convection in this setting.
To determine the impact of dimensionality on large-scale stellar convection, the convective dynamics in two- and three-dimensions must be compared directly.



Direct comparisons of two- and three-dimensional convection have been made for Rayleigh-B\'enard convection \citep{schmalzl2004validity,van2013comparison,goluskin2016penetrative}.  
Unlike stellar convection, Rayleigh-B\'enard convection is defined by a convecting fluid constrained between two solid plates that produce a temperature gradient, and is studied in a controlled laboratory setting.   
Rayleigh-B\'enard convection simulations are performed as direct numerical simulations that solve the Boussinesq convection equations; therefore the small scales of turbulence are expected to be sufficiently resolved, at least in so far as they impact the large-scale properties of the convection, including the thickness of the boundary layers at the solid plates.
Observed differences between two- and three-dimensional Rayleigh-B\'enard convection are dependent on Prandtl number \citep{schmalzl2004validity,van2013comparison,goluskin2016penetrative}.  The explanation for this dependence is that the Prandtl number affects the dominant shapes of plumes.  At high Prandtl number the shape of plumes is closely comparable in two and three dimensions.  Thus for Prandtl number greater than one,  simulations in two- and three-dimensions produce similar Nusselt and Reynolds numbers, and these numbers converge with increasing Prandtl number.  Two- and three-dimensional simulations also produce similar thermal profiles with depth in this large Prandtl number regime.
In contrast, in the low Prandtl number regime that is expected in the stellar interior, two- and three-dimensional convection results diverge.  These studies have found that the Reynolds number is larger for two-dimensional simulations than for three-dimensional simulations.  Furthermore, the Nusselt number is similar for two- and three-dimensional simulations, independent of the Prandtl number, as long as the Rayleigh number is sufficiently high \citep[e.g. $\mathsf{Ra} \sim 10^8$ in][]{van2013comparison}.  These intriguing results point to a need for better understanding of the dimensional properties of stellar convection, a physically more complicated and less controlled setting than Rayleigh-B\'enard convection.

\hspace{.5mm}


Direct comparisons of two- and three-dimensional convection have also been made for atmospheric convection.  Atmospheric models use large-eddy simulations (LES) and most commonly solve the equations for anelastic convection, although Boussinesq, or compressible equations have also been used.  LES simulations model the effect of small-scale turbulence using a sub-grid scale model, and in the case of atmospheric convection commonly include microphysics relevant to clouds.  Several studies of dimensionality have been made for atmospheric convection  \citep{moeng1996simulation,moeng2004investigating,phillips2006cloud,petch2008differences}.  In those studies, higher vertical velocities in three-dimensional simulations have been reported than in two-dimensional simulations.  This ordering is opposite to the commonly found results for stars \citep{muthsam1995numerical,meakin2007turbulent,arnett2011toward}.  These atmospheric convection studies also report a smaller depth of entrainment and lower level of mixing in two-dimensional simulations.    They find that differences between two- and three- dimensional simulations are sensitive to boundary conditions such as convection over land or water.    In addition, when a two-dimensional LES is calibrated to data, including tuning of the sub-grid scale method, results are encouragingly similar to three-dimensional simulations of atmospheric convection \citep{moeng1996simulation,moeng2004investigating}.\footnote{It is relevant to the stellar 321D-link for stellar evolution to observe that the field of atmospheric modeling also pursues more accurate 1D models of convection.  In climate modeling these are called convection parameterizations or superparameterizations \citep{randall2003breaking,tao2009multiscale,majda2014new,piriou2007approach}. 
} 

\hspace{.5mm}




Stellar convection differs from these two other convection settings in key ways.  For stellar convection, the simulation volume is spherical and the size of convective flow structures may be independent of a simulation's aspect ratio.   The fluid is internally heated and radiates energy, and the convection interacts with stratified density and a temperature gradients that extend over a significant portion of the stellar radius.  This stratification indicates that for most stars the fundamental parameters will cover a range of low Prandtl number, high Rayleigh number, high Reynolds number, and high Nusselt number regimes; however these fundamental parameters vary significantly throughout the stellar radius, and can vary in a different way for different types of stars.
   The treatment of boundaries on a convecting layer in a star is also considerably different from either Rayleigh-B\'enard convection or atmospheric convection; a stellar convection zone is typically bordered by layers of convectively stable stellar material, so that  characterizing convective overshooting and penetration is important to a full description of stellar convection.   In addition, the different physical models of Boussinesq convection, anelastic convection, or fully compressible convection may produce different results when comparing 2D and 3D simulations.  Each of these points mean that although previous studies may inform our expectations, they cannot be used to directly predict the different properties of two- and three-dimensional compressible stellar convection.  It is useful to perform additional comparisons targeted toward global simulations performed as ILES of compressible stellar convection.

This work is structured as follows.  In Section \ref{secmodel} we briefly summarize the young sun model that has been completely described in \citet{prattspherical}, and discuss the numerical framework of our simulations.  In Section \ref{secresultsid} we discuss statistical results related to the flow field to allow us to compare two-dimensional and three-dimensional stellar dynamics.  This section expands on the common result that 2D velocities are higher than 3D velocities \citep{muthsam1995numerical,meakin2007turbulent,arnett2011toward}.
  In Section~\ref{secpenetration}, we present calculations of the penetration depth and a one-dimensional diffusion coefficient enhanced by convective mixing, as proposed by \citet{pratt2017extreme}.  We compare this enhanced diffusion coefficient for two- and three-dimensional simulations.  In Section \ref{secconc} we discuss the implications of these results for multi-dimensional explorations related to the 321D link, and for further studies of convective penetration in stars.

\section{Simulations \label{secmodel}}

In this work we use the numerical set-up for stellar convection described and examined in \citet{prattspherical}; we refer to that earlier work for the full details of the star beyond the brief summary here.
We perform two-dimensional and three-dimensional ILES of a prototypical low-mass pre-main sequence star, called the young sun, using the \music code. 
The young sun we examine weighs one solar mass and has homogeneous chemical composition, consistent with models of the sun at an early evolutionary state.  The radial profiles of density and temperature for the young sun are typical for a pre-main sequence star that is no longer accreting and is gradually contracting.  A central radiative zone exists below the large convection zone; the young sun is convectively unstable over $1.2 \cdot 10^{11}$ cm of the total radius of $2.13 \cdot 10^{11}$ cm.  This large convective envelope allows us to study deep stellar convection, far from the physically complicated near-surface layers. 
Our simulations of the young sun in this work only take convection into account; the possibility of studying additional physical effects such as rotation, a tachocline, chemical mixing, and magnetic fields are omitted from the current study which focuses on expanding our convection results from two-dimensional stellar convection to three-dimensions.  We study the dynamics of convection in this realistic stratification for a star; a study of wave dynamics is beyond the scope of this work.

\vspace{2mm}The \music code solves the inviscid compressible hydrodynamic equations for density $\rho$, momentum $\rho \vec{u}$, and internal energy $\rho e$:
\begin{eqnarray} \label{densityeq}
\frac{\partial}{\partial t} \rho &=& -\nabla \cdot (\rho \vec{u})~,
\\ \label{momeq}
\frac{\partial}{\partial t} \rho \vec{u} &=& -\nabla \cdot (\rho \vec{u} \vec{u}) - \nabla p + \rho \vec{g} ~,
\\ \label{ieneq}
\frac{\partial}{\partial t} \rho e &=& -\nabla \cdot (\rho e\vec{u}) +p \nabla \cdot \vec{u} + \nabla \cdot (\chi \nabla T) .
\end{eqnarray}
using a finite volume method, a MUSCL method \citep{thornber2008improved} of interpolation, and a van Leer flux limiter \citep[as described in][]{van1974towards,roe1986characteristic,leveque2006computational}.   For two-dimensional simulations, the finite volume method assumes azimuthal symmetry.  Time integration in the \music code is fully implicit, and uses a Jacobian free Newton-Krylov (JFNK) solver \citep{knoll2004jacobian} with physics-based preconditioning \citep{maximepaper}.  We use an adaptive time step, which is constrained identically for two- and three-dimensional simulations.


\music simulations are designed to contribute to the 321D link \citep{david20143d, arnett2009turbulent}; one aspect of this is the use of an equation of state and realistic opacities that are standardly used in one-dimensional stellar evolution calculations. Opacities are interpolated from the OPAL \citep{iglesias1996updated} and \citet{ferguson2005low} tables, which cover a temperature range suitable for the description of the entire structure of a low-mass star like the young sun. The compressible hydrodynamic equations \eqref{densityeq}-\eqref{ieneq} are closed by determining the gas pressure $p(\rho,e)$ and temperature $T(\rho,e)$ from a tabulated equation of state for a solar composition mixture.
This equation of state accounts for partial ionization of atomic species by solving the Saha equation, and neglects partial degeneracy of electrons; it is suitable for the description of a solar model at a young age. The initial state for \music simulations is produced using data extracted from a one-dimensional model calculated from the Lyon stellar evolution code \citep{baraffe1991evolution, baraffe1997evolutionary,baraffe1998evolutionary}, which uses the same opacities and equation of state as \music.
 In eq. \eqref{momeq}, $\vec{g}$ is the gravitational acceleration, a spherically-symmetric vector consistent with that used in the Lyon stellar evolution code, and not evolved by our simulations.  

\subsection{Spherical-shell geometry and boundary conditions \label{secbc}}
    
The compressible hydrodynamic equations \eqref{densityeq}-\eqref{ieneq} are solved in a spherical shell using spherical coordinates:  radius $r$, and angular variables $\theta$ and $\phi$ (in 3D).  Our simulations of compressible hydrodynamic convection are summarized in Table~\ref{simsuma}.  In this table, the inner and outer radius of the spherical shell for each simulation is noted, and the radial and angular grid spacings are specified.   The two-dimensional $r$ -- $\theta$ simulation volume and grid is identical for each pair of simulations considered.   This is an important detail for an implicit large-eddy simulation; grid spacing is directly related to the effective numerical viscosity, and the fluid properties in the convection zones should be comparable.
  Our simulations have sufficient radial resolution to produce a characteristic radial profile for velocity in 2D; this has been compared with 2D simulations that use up to 2624 radial grid points.  Convergence toward a velocity profile is observed as resolution is increased.   The three-dimensional simulation \emph{wide3D} has a grid of $r \times \theta \times \phi = 320 \times 256 \times 256$.
   This grid size is selected for this study, so that in 3D a total simulation time of sufficient length can be produced for the comparison of statistical quantities with 2D simulations.  Each 3D simulation has been relaxed efficiently into a three-dimensional steady-state flow beginning from a realistic velocity perturbation extracted from the accompanying 2D simulation.  All data presented here is produced during steady-state convection, a period where the time-averaged value of the total kinetic energy is well-defined and not changing in time.
  
To compare the statistics of convection and particularly of convective penetration in a large convection zone to those in a smaller convection zone with the same local stratification, we have produced two variations of simulations described in Table~\ref{simsuma}: (1) the \emph{wide} and \emph{deep} simulations simulate the full convection zone of the young sun, and (2) the \emph{short-a} and \emph{short-b} simulations simulate a truncated convection zone.  The simulations in a truncated convection zone are conceptually similar to early ``box-in-a-star'' studies of convective penetration that include a limited local region around the convective boundary \citep[e.g.][and references therein]{brummell2002penetration}.  

 
\begin{table*}
\begin{center}
\caption{Parameters for compressible hydrodynamic simulations of the young sun.
 \label{simsuma}
 }
\begin{tabular}{lccccccccccccccccccccccccccc}
                      & dimensions & $R_{\mathsf{i}}/R$  & $R_{\mathsf{o}}/R$ & $\Delta r/R$  & $\phi$ (\degree) & $\Delta \theta$ (\degree)  &  $\tau_{\mathsf{conv}}$($10^6$s) & time span ($\tau_{\mathsf{conv}}$) 
\\ \hline
wide3D               & 3D & 0.21         & 1.00  & $28 \cdot 10^{-4}$   &  140 & 0.55 &   $ 0.70 \pm 0.04 $  & 4.09  
\\ \hline
deep3D               & 3D & 0.10         & 1.00  & $28 \cdot 10^{-4}$ &  30  & 0.55 &   $ 0.77 \pm 0.03 $  & 4.05  
\\ \hline
short3Da              & 3D & 0.31         & 0.67   & $28 \cdot 10^{-4}$ & 140   & 0.55 &   $ 4.47 \pm 1.67 $  & 3.36  
\\ \hline
short3Db               & 3D & 0.31         & 0.67  & $28 \cdot 10^{-4}$  &  140   &1.10 &   $ 5.71 \pm 1.71 $  & 8.30  
\\ \hline  \hline
wide2D                 & 2D & 0.21         & 1.00  & $28 \cdot 10^{-4}$  &    0    & 0.55 &  $ 0.79 \pm 0.07 $  & 104      
\\ \hline
deep2D               & 2D & 0.10         & 1.00   & $28 \cdot 10^{-4}$ &    0     &0.55 &   $ 0.94 \pm 0.13 $  & 284  
\\ \hline
short2Da               & 2D & 0.31         & 0.67  & $28 \cdot 10^{-4}$ &    0     &0.55 &   $ 2.20 \pm 0.29 $  & 121  
\\ \hline
short2Db                 & 2D  & 0.31        & 0.67  & $28 \cdot 10^{-4}$   &    0    &1.10 &  $ 2.57 \pm 0.32 $& 174   
\\ \hline \hline
\end{tabular}
\tablefoot{The inner $R_{\mathsf{i}}$ and outer $R_{\mathsf{o}}$ radius of the spherical shell, and the radial grid spacing $\Delta r$ in the convection zone are given in units of the total radius of the young sun $R$.   The angular extent of the simulation in the 3rd direction is $\phi$ and the grid spacing in both angular directions is $\Delta \theta$.  The average global convective turnover time $\tau_{\mathsf{conv}}$ is provided, and the total time span for each simulation is given in units of the convective turnover time. }
\end{center}
\end{table*}
 
In \citet{prattspherical}, we studied the placement of boundaries and the choice of boundary conditions; these choices were found to affect the physical outcome of our hydrodynamic simulation.  In this work, we examine simulations with two variations on sets of boundary conditions that each maintain the original radial profiles of density and temperature of the one-dimensional stellar evolution model.  For the \emph{wide} and \emph{deep} simulations, which include the full stellar radius, we allow the surface to radiate energy with the local surface temperature.  In this case the energy flux varies as $\sigma T_{\mathsf{s}}^4$ where $\sigma$ is the Stefan-Boltzmann constant and $T_{\mathsf{s}}(\theta,t)$ is the temperature along the surface.  This boundary condition can only be effectively used when the near-surface layers are included in the simulation volume and the temperature gradient near the surface is sufficiently resolved; otherwise it results in artificially high cooling rates.  For the  \emph{short-a} and \emph{short-b} simulations, which do not include the full stellar radius, we hold the energy flux and luminosity constant on the outer radial boundary, at values established from the one-dimensional stellar evolution calculation.  For an examination of how these boundary conditions affect the dynamics, we refer to \citet{prattspherical}.

Aside from this surface boundary condition, the conditions set at other boundaries are the same across all simulations. Each simulation volume begins at $20 \degree$ from the north pole, and ends $20 \degree$ before the south pole.  In the three-dimensional simulations, the additional angular extent in $\phi$ is given in Table~\ref{simsuma}.  We impose periodicity on all physical quantities at the boundaries in $\theta$ and $\phi$.   In velocity, we impose non-penetrative and stress-free boundary conditions in the radial directions.  The energy flux and luminosity are held constant at the inner radial boundary, at the value of the energy flux at that radius in the one-dimensional stellar evolution calculation.   
On the inner radial boundary of the spherical shell, we impose a constant radial derivative on the density.  At the outer radial boundary we apply a hydrostatic equilibrium boundary condition on the density that maintains hydrostatic equilibrium by assuming constant internal energy and constant radial acceleration due to gravity in the boundary cells \citep{hsegrimm2015realistic}.  These boundary conditions allow us to closely match the stratification of density at the boundaries of our simulation to the structure of the young sun \citep[see][]{prattspherical}.

\subsection{Fundamental parameters}

It has been established in DNS of Rayleigh--B\'enard convection \citep{schmalzl2004validity,van2013comparison,goluskin2016penetrative} that the value of the Prandtl number affects a comparison of two- and three-dimensional dynamics.
In DNS of convective overshooting in a box, \citet{brummell2002penetration} find that the Peclet number plays a significant role.
To establish these dependencies, these DNS studies perform a range of simulations for carefully controlled, different values of the fundamental parameters.
 Unlike such studies of DNS, the present work examines global ILES of a single stellar structure. 
  We do not seek to reproduce the results of these earlier works using global simulations using the ILES simulation framework; such a study would be an enormous undertaking in the context of the realistic stellar structure models which we are studying.

We also do not seek to compare directly with the results of earlier DNS studies.  Because the grid spacing is large in global ILES simulations, the effective values of the Prandtl number and Peclet number produced are inevitably more moderate than the values possible for a DNS simulation of comparable computational size.
  In any ILES simulation framework the values of the viscosity and thermal diffusivity are not explicitly specified.  In spherical stellar simulations they vary throughout the radius of a star, dependent both on grid structure and properties of the stellar model.  An estimation of these parameters in our simulations would be crude in comparison to a DNS. 
Moreover, in the young sun the velocities and their length scales at a given radius have a wide variation linked to the particular structure of this star.   The largest velocity convection rolls in the young sun can be associated with multiple characteristic length scales at a given depth, contributing toward a more general ambiguity.  This observation is in agreement with the ideas of non-local convection in a large convection zone.  Thus an interpretation of our results with respect to the fundamental parameters of the flow is not simple, and will not be pursued further here.  We note that these properties likely do not hold for all stellar structures.



\section{Results: average dynamics \label{secresultsid}}

The stratification of temperature and density in the young sun change on a thermal time scale that is much longer than any of the simulations considered in this work.  The radial profiles of these quantities are initially identical in our two- and three-dimensional simulations and do not deviate significantly from their initial values during our simulations.  Therefore we compare quantities derived from the velocity dynamics, considered over several convective turnover times during stationary steady-state convection.


\subsection{Comparison of velocities \label{secresultsbasic}}

\begin{figure*}
\begin{center}
\resizebox{3.5in}{!}{\includegraphics{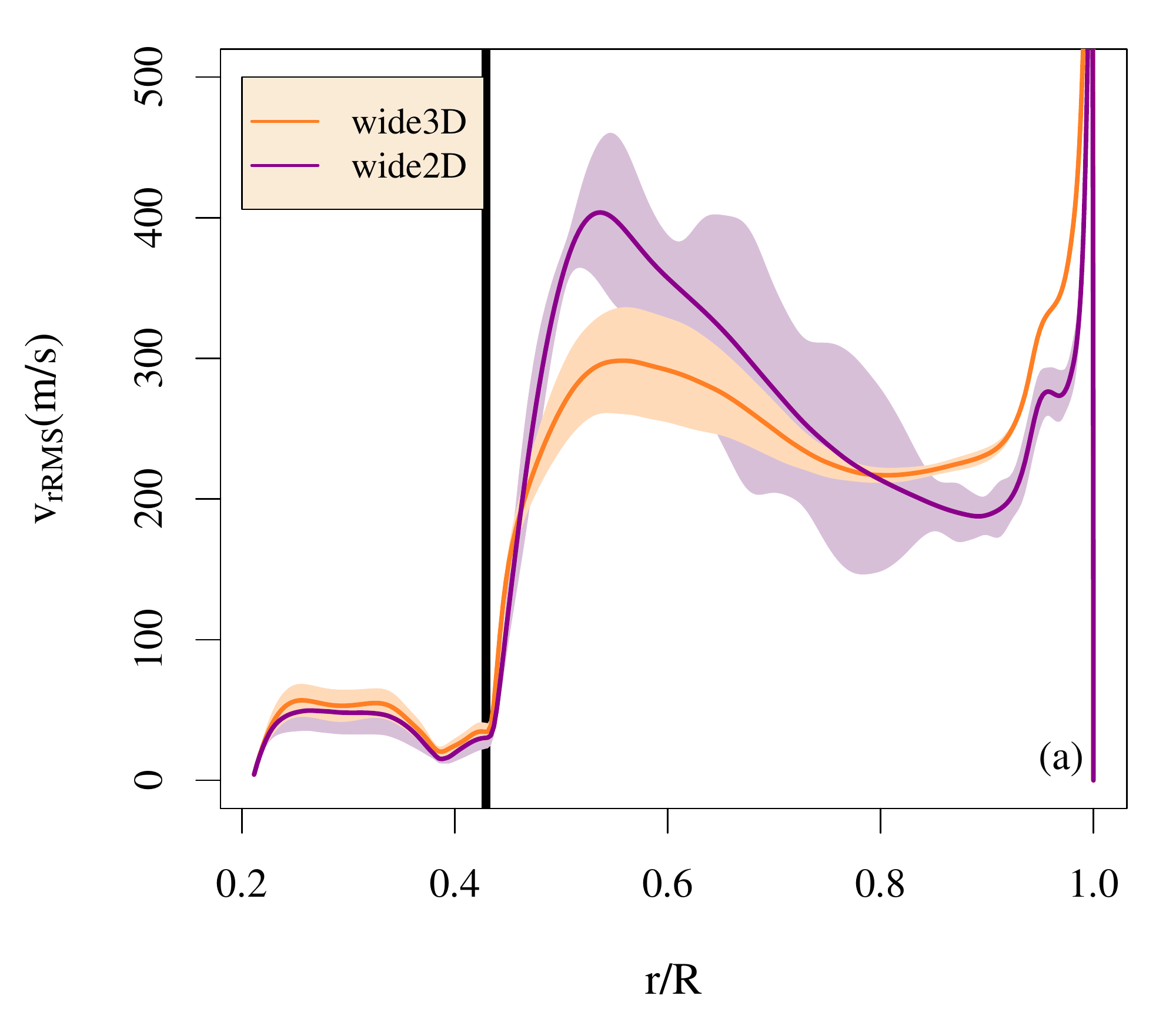}}\resizebox{3.5in}{!}{\includegraphics{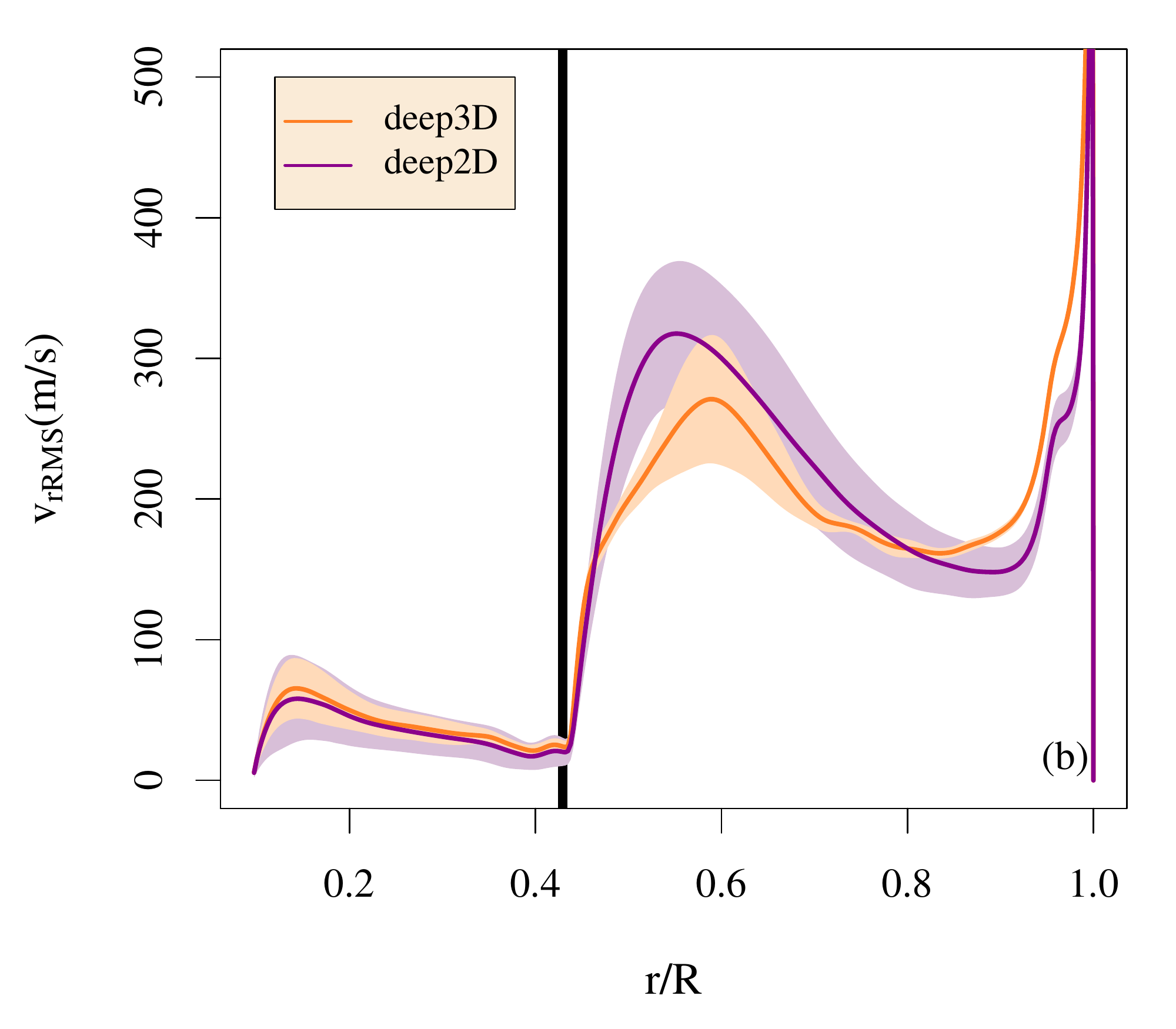}}
\resizebox{3.5in}{!}{\includegraphics{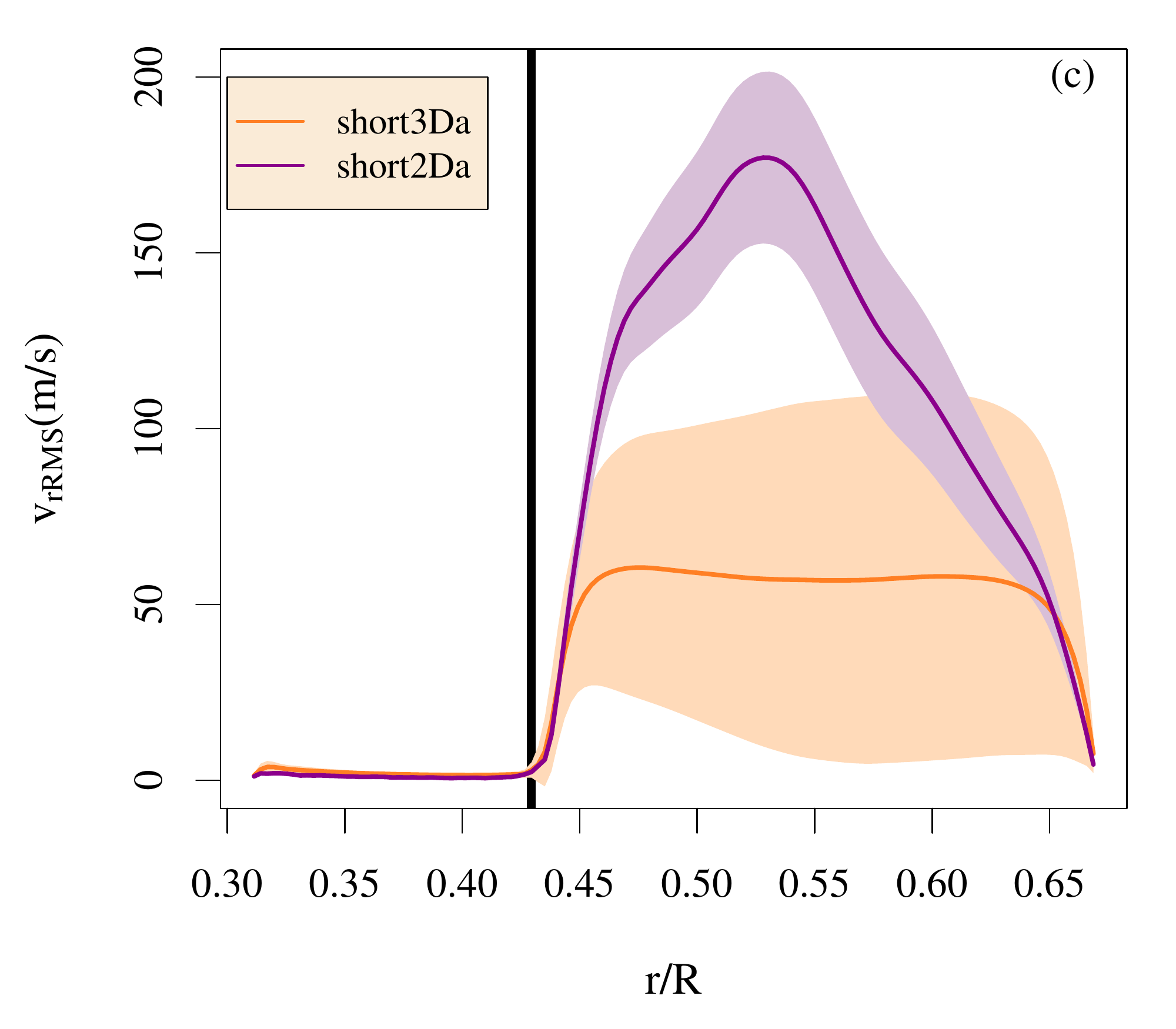}}\resizebox{3.5in}{!}{\includegraphics{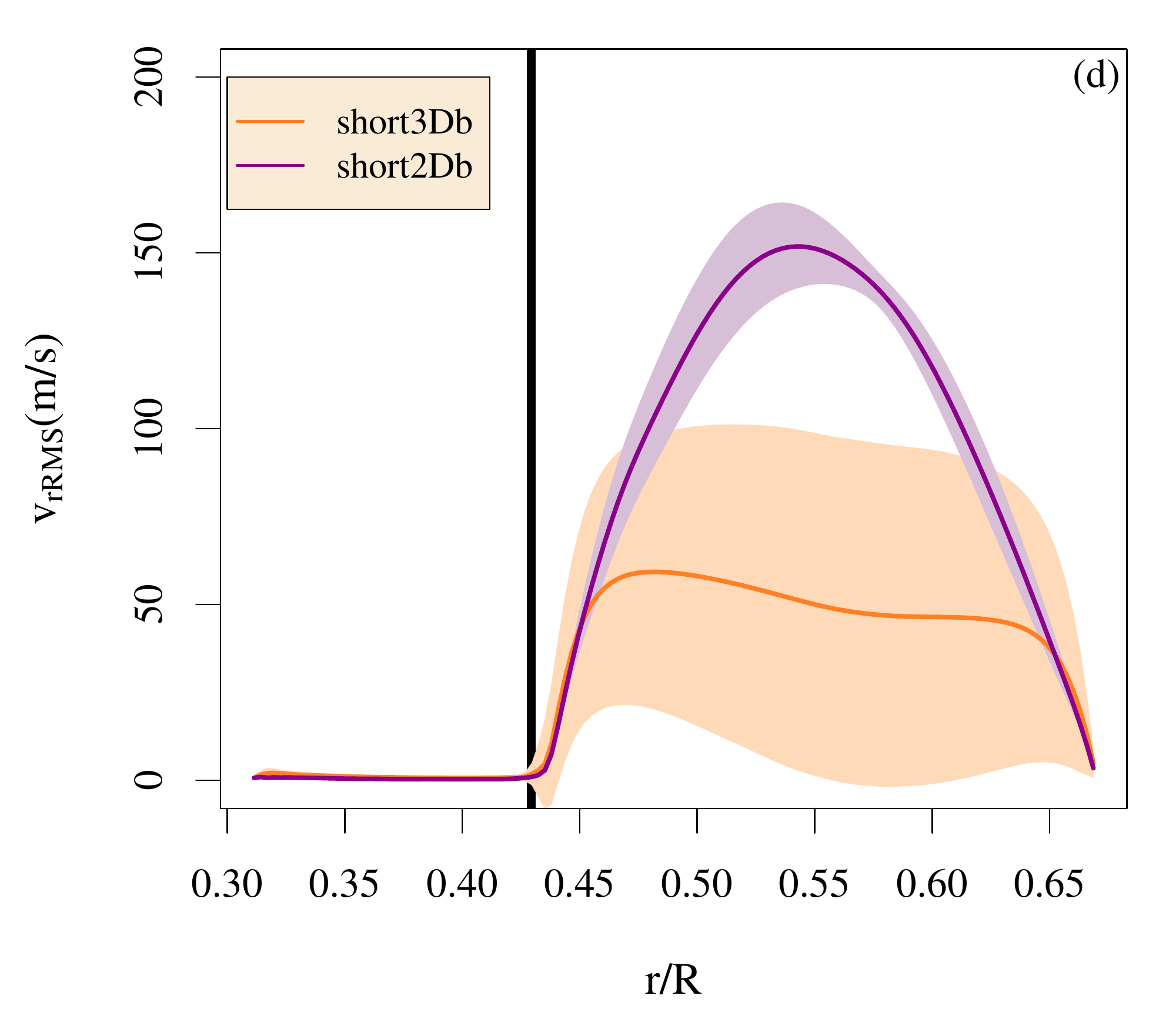}}
\caption{Root-mean-square radial velocity (a) in simulations \emph{wide2D} and \emph{wide3D}, (b) in simulations \emph{deep2D} and \emph{deep3D}, (c) in simulations \emph{short2Da} and \emph{short3Da},  (d) in simulations \emph{short2Db} and \emph{short3Db}.  Shaded areas indicate one standard deviation above and below the time-averaged line.   The heavy vertical line indicates the bottom of the convection zone determined by the Schwarzschild criterion.
\label{figvradrms}}
\end{center}
\end{figure*}

Studies of stellar convection have previously found that two-dimensional fluid simulations result in higher velocities than three-dimensional simulations, and we confirm that result.  
 Fig.~\ref{figvradrms} shows time-averaged profiles of the root-mean-square (RMS) of the radial velocity for the four pairs of two-dimensional and three-dimensional simulations that we performed for this study.    The profiles of RMS radial velocities are averaged in the horizontal directions for the ``mean'' aspect of the RMS, and then averaged in time.   The shaded areas indicate one standard deviation above and below the time-averaged line to supply information on the variation of these RMS velocities.  In the \emph{deep} and \emph{wide} simulations the variation in 2D and 3D radial velocities is comparable, with slightly more variation in 2D.  The variation in 3D radial velocities is significantly larger than in 2D for the \emph{short-a} and \emph{short-b} simulations which are performed in a truncated convection zone with constant energy flux imposed on the outer radial boundary.  Thus in the radial velocity field, boundary conditions and the choice of spherical shell have a demonstrable effect on the differences between 2D and 3D flow variation.

The RMS radial velocities have similarly shaped profiles for each pair of 2D and 3D simulations, characterized by larger values in the middle of the convection zone.   The \emph{short-b} simulations have lower velocities than the \emph{short-a} simulations, consistent with expectations that the lower grid resolution in the angular directions of the \emph{short-b} simulations creates higher numerical dissipation.  A comparison between the \emph{wide} simulations and the \emph{short-a} simulations shows that the \emph{short} simulations have slightly less than half of the RMS radial velocity.  Truncating the simulation volume in the upper convection zone and fixing the energy flux at that boundary results in lower RMS velocities than in simulations that reach to the surface.  For the young sun, a range of tests \citep{prattspherical} show that both the truncation of the convection zone, the removal of outer layers that have a different stratification, and the use of a different boundary condition contribute to this difference.  This supports the view that both full-star simulations and realistic boundary conditions are important steps toward achieving more realistic stellar flows.


The velocities at the convective boundary are of particular interest for 
models of convective penetration.  In the narrow region above and directly surrounding the convective boundary, between approximately $0.3 \leq r/R \leq 0.45$, the RMS radial velocities are nearly identical between each pair of 2D and 3D simulations, and have a similar shaped drop toward the radiative zone.  The velocities immediately above the radiative zone are less for the \emph{short} simulations than the  \emph{wide} or \emph{deep} simulations.  Full-star simulations and realistic boundary conditions are therefore particularly important to achieving realistic simulations of convective plumes overshooting in the deep stellar interior.
 
The RMS of the velocities in the $\theta$ direction, and the RMS of the full velocity vector both follow a similar trend where 2D velocities are larger than 3D velocities.  The RMS of the velocity vector averaged over the convection zone is 30\% larger in 2D for the \emph{wide} simulations, 18\% larger in 2D for the \emph{deep} simulations, 84\% larger in 2D for the \emph{short-a} simulations, and 109\% larger in 2D for the \emph{short-b} simulations.   Thus truncating the convection zone and using a fixed energy flux, as is done in the \emph{short-a/short-b} simulations,  exaggerates the differences between 2D and 3D simulations in the young sun.  We also find that the velocity ratio of average $v_{r,\mathsf{RMS}}/v_{\theta,\mathsf{RMS}}$ is lower in each of our 2D simulations than in 3D simulations.   In a broad sense, the ratio of radial to angular velocity implies that the flow has a different topological structure. 

We can compare the results in Fig.~\ref{figvradrms} with Figure~6 and Figure~12 of \citet{meakin2007turbulent}.  That work compares velocities from 2D and 3D simulations of a star with convectively stable oxygen shell burning, and a star with a convecting core.
The grid size for their simulations is $400 \times 100 \times 100$, which is similar to our simulations, although the resolution of the flows in these physically different stars is difficult to compare.\footnote{The convective velocities produced in the simulations of \citet{meakin2007turbulent} are larger than the velocities in the large convection zone of our young sun, possibly linked to a different non-linear stratification in temperature and density for these stars, and a different interaction between the convection and stratification.  This points to the need for clear physical definitions of resolution in ILES so that results can be compared.}  In their 2D simulations the peak velocities are approximately 2 -- 5 times larger than in their 3D simulations, a result close to our \emph{short-a} and \emph{short-b} results.   This reinforces the idea that differences between 2D and 3D convection depend on boundary conditions, resolution\footnote{The resolution may be linked to variation of fundamental parameters of the flow; those dependencies have been examined in studies of DNS, and are not explored directly in this work.}, the size and structure of the convection zone, and possibly also the presence and structure of neighboring stable layers.  These differences thus need to be studied for different global stellar models; such a study is underway to analyze the effect of nearby stable layers on convection in the current sun as an additional point of comparison \citep[][in prep.]{dimitarinprep}.

\begin{figure*}
\begin{center}
\resizebox{2.2in}{!}{\includegraphics{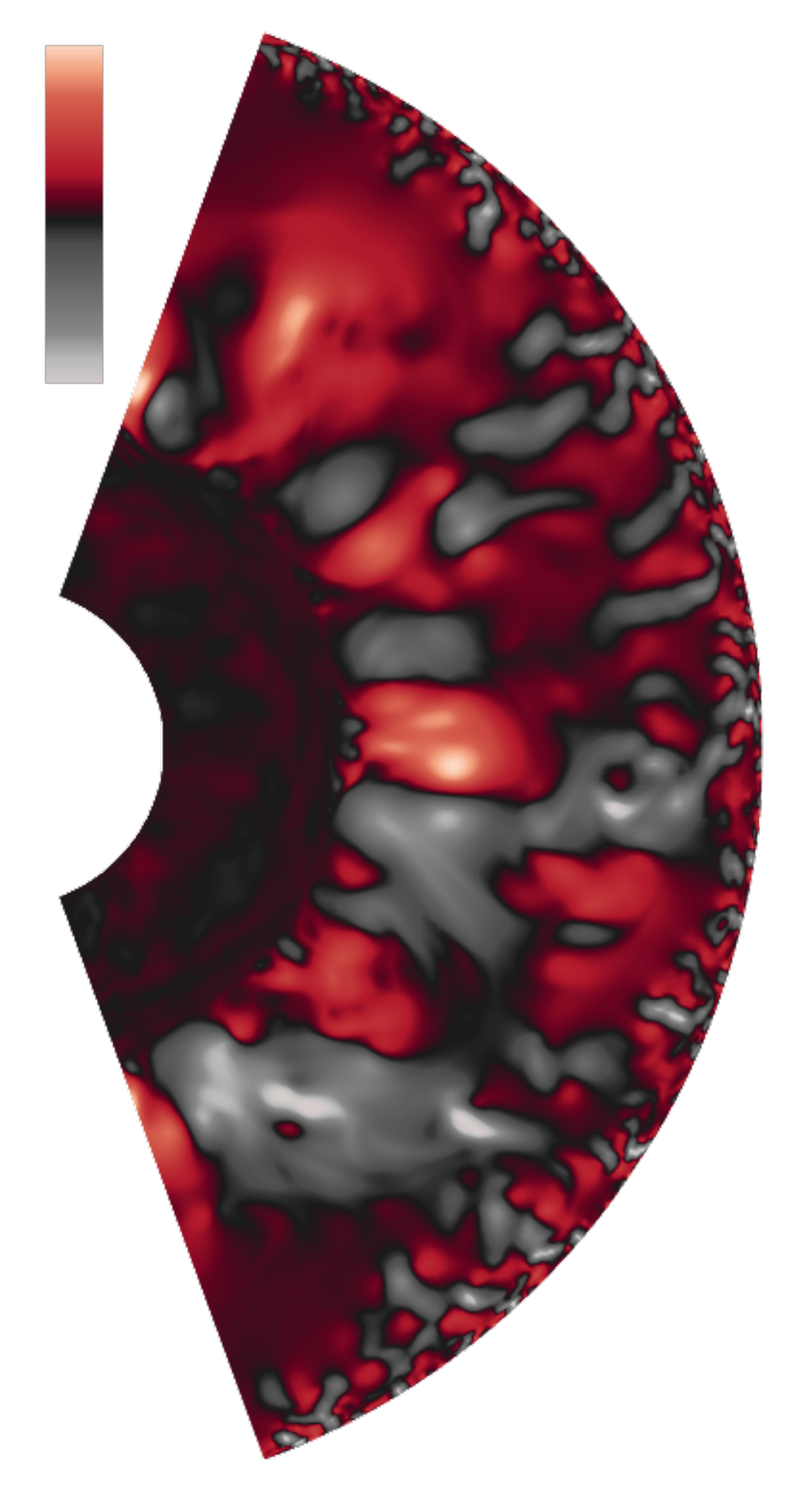}}\resizebox{2.15in}{!}{\includegraphics{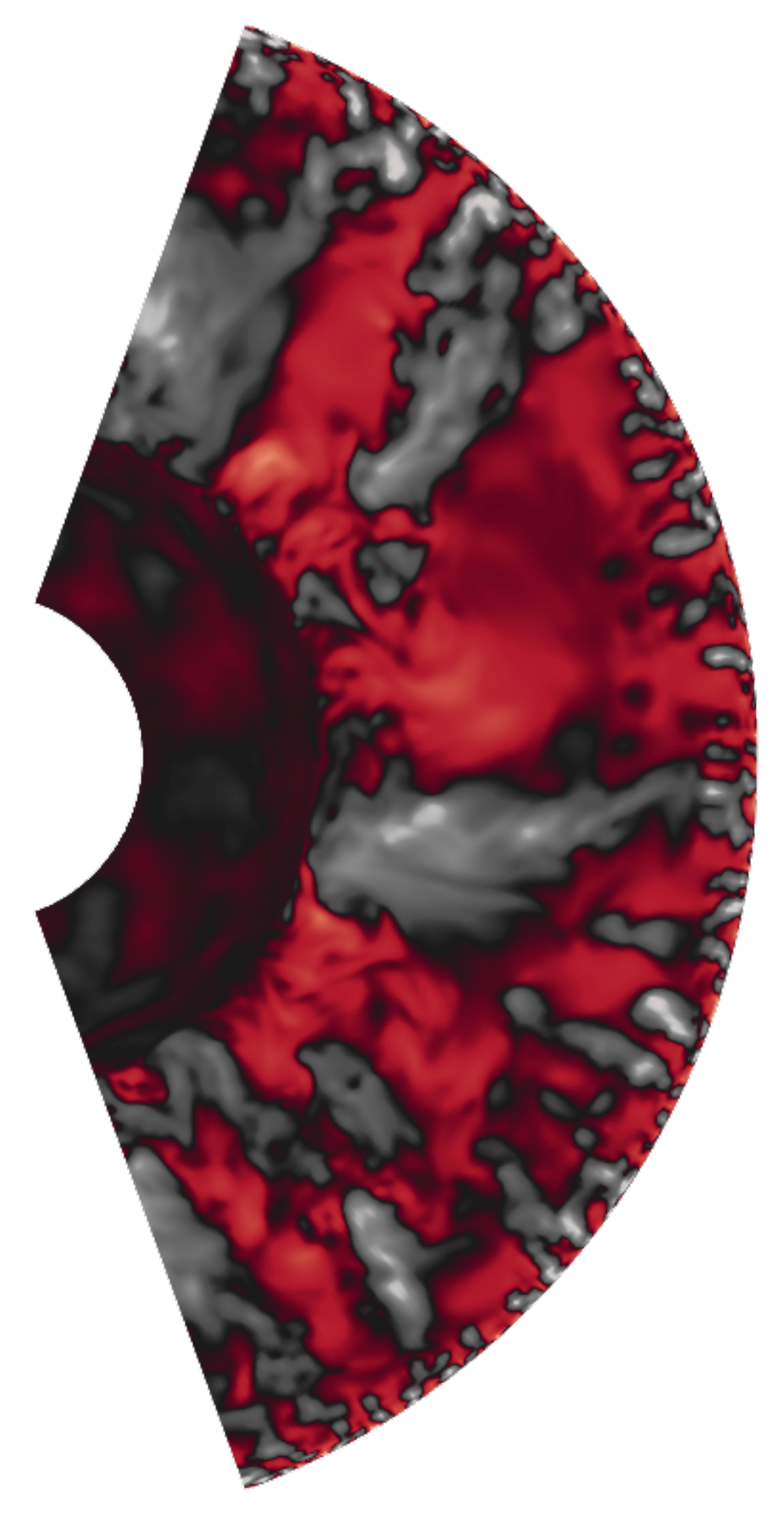}}
\caption{Typical snapshot of the radial velocity in simulation \emph{wide2D} (left), and in a two-dimensional cut of simulation \emph{wide3D} (right).  Color scales are identical.   Red indicates an outflow, while grey indicates an inflow.
\label{figradvelviz}}
\end{center}
\end{figure*}

Flow visualizations provide additional information about the structure of the velocity field. A typical snapshot of the radial velocities in simulations \emph{wide2D} and \emph{wide3D}, with identical color scales, is shown in Fig.~\ref{figradvelviz}.   Both 2D and 3D simulations have similar size small-scale convective structures in the near-surface layers of the simulation.  This layer of small-scale near-surface convection also has a similar radial width in 2D and 3D.  Beneath the near-surface layers, larger scale radial flows develop in both simulations.  The size of these large-scale convective flows is comparable in 2D and 3D.  In the 3D simulation, these large-scale radial velocity structures have greater ``roughness'', or small-scale irregularities of the flow.  The radiative zones of these 2D and 3D simulations look indistinguishable.

\subsection{Comparison of vorticity  \label{secresultsvort}}

\begin{figure*}
\begin{center}
\resizebox{2.2in}{!}{\includegraphics{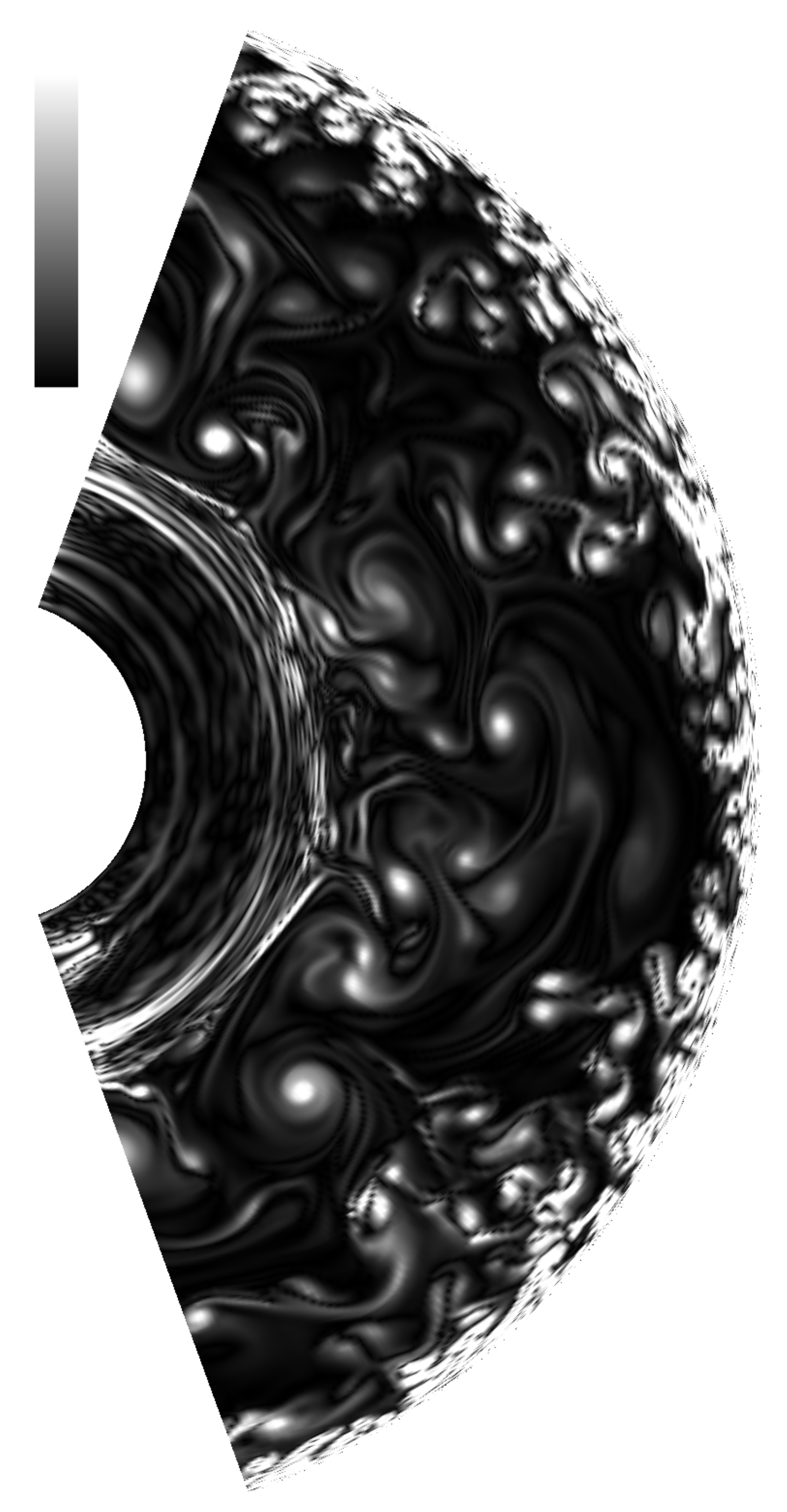}}\resizebox{2.27in}{!}{\includegraphics{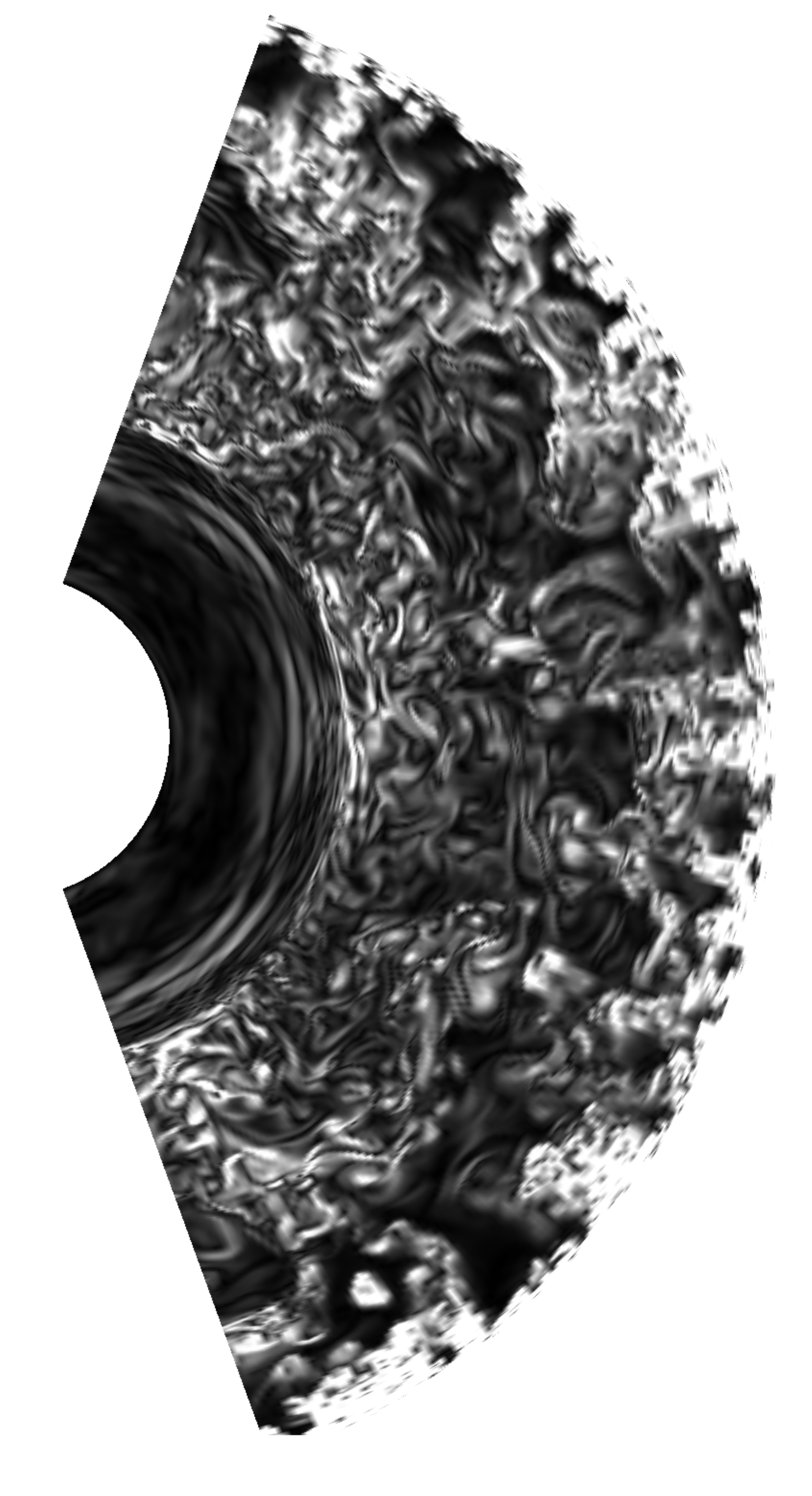}}
\caption{Typical snapshot of the vorticity magnitude in simulation \emph{wide2D} (left), and in a two-dimensional cut of simulation \emph{wide3D} (right).  Color scales are identical.  Lighter color indicates higher vorticity magnitude.
\label{figvortviz}}
\end{center}
\end{figure*}

 To further probe the different topological structuring in 2D and 3D flows, we examine the vorticity.   A visualization of a typical snapshot of vorticity magnitude in simulations  \emph{wide2D} and  \emph{wide3D}, with identical color scales, is shown in Fig.~\ref{figvortviz}.  The vorticity of the smaller scale convective flows in the near-surface layers, near the outer radius of the simulations, appears to be similar in 2D and 3D.  In the middle and lower layers of the convection zone, the 2D simulation has much larger, smoother, more coherent structures in vorticity than the 3D simulation.  The salient features of these visualizations are consistent with those produced in the recent work of \citet{zingale2015comparisons}.

\begin{figure*}
\begin{center}
\resizebox{3.5in}{!}{\includegraphics{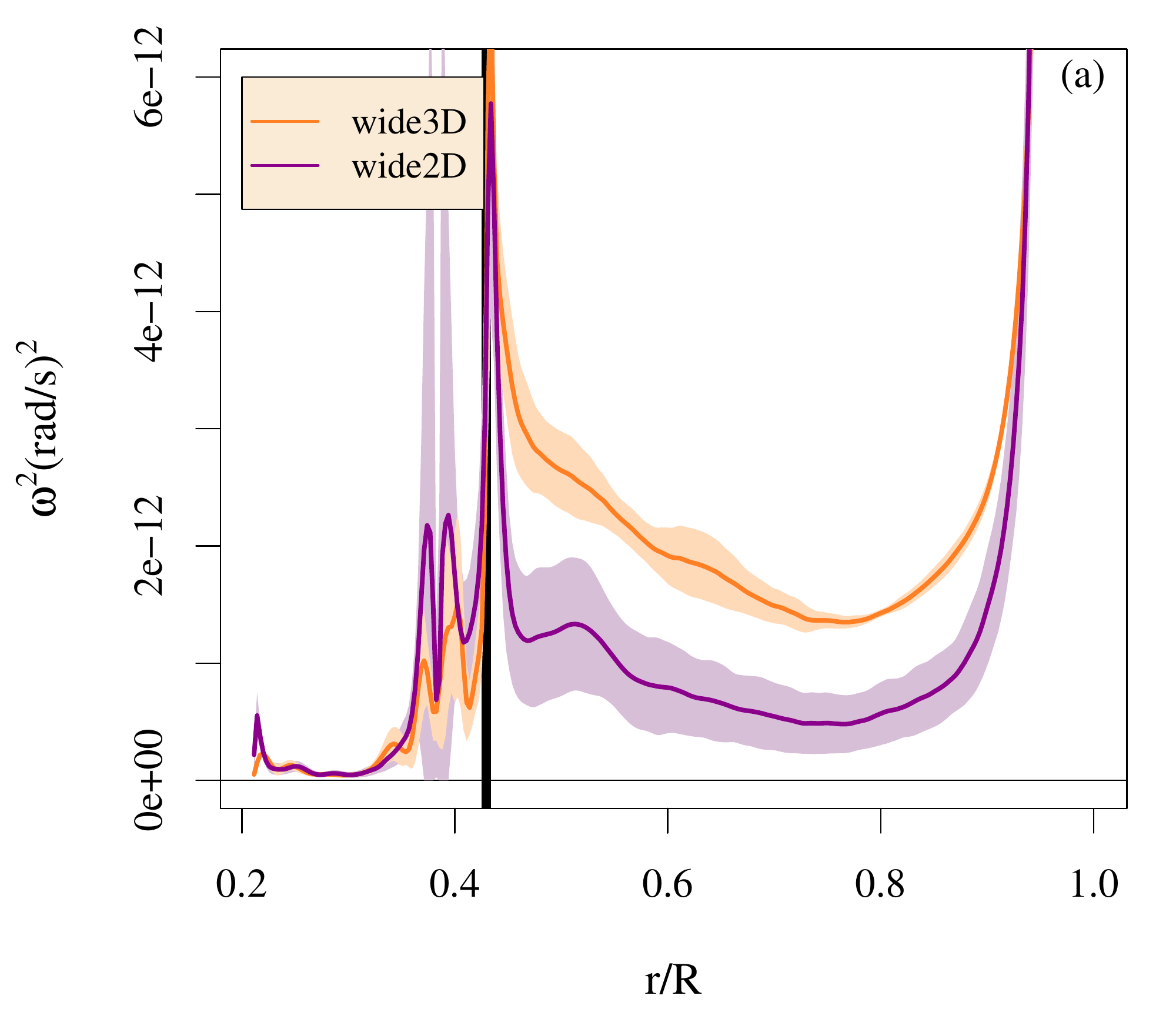}}\resizebox{3.5in}{!}{\includegraphics{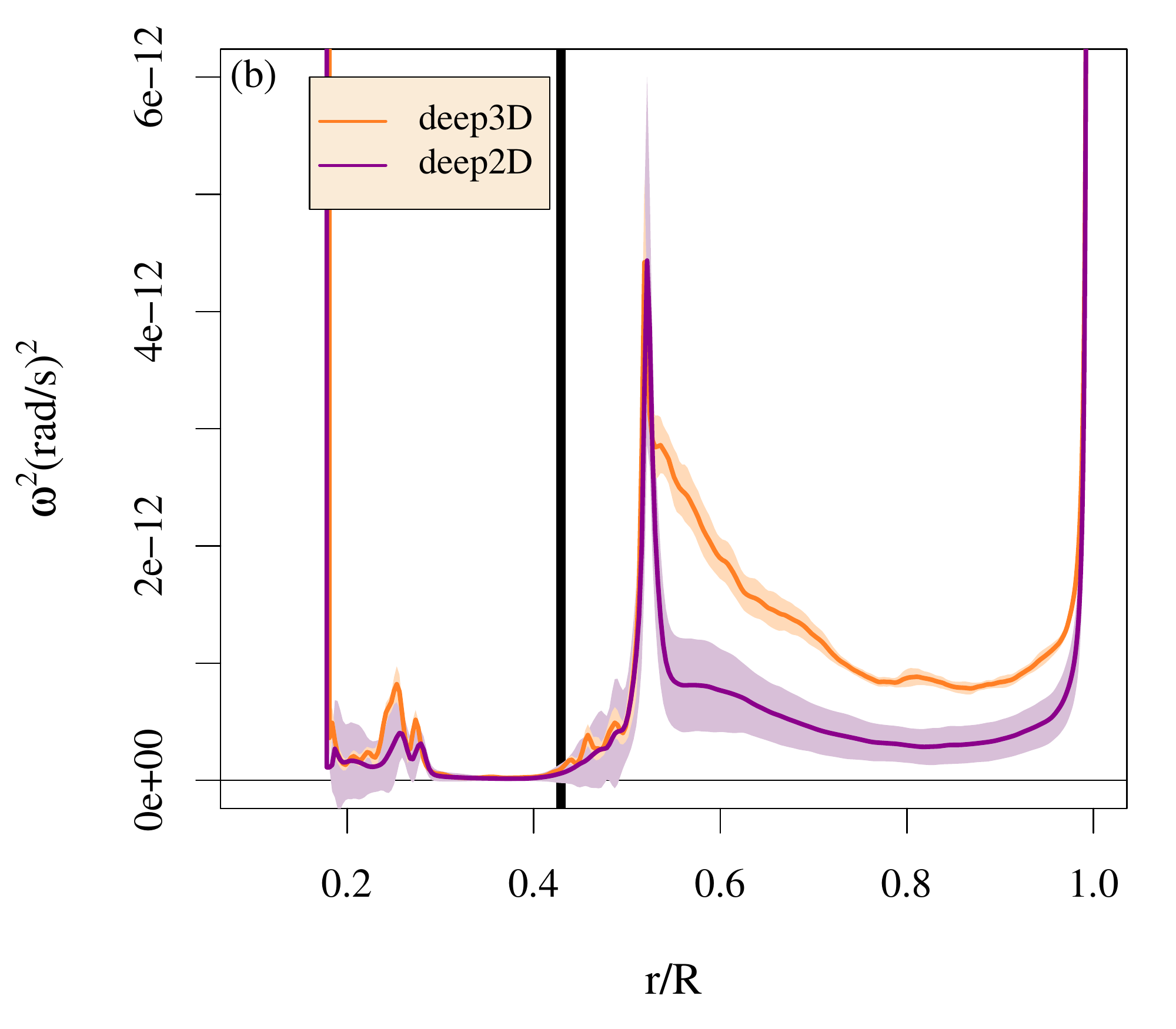}}
\resizebox{3.5in}{!}{\includegraphics{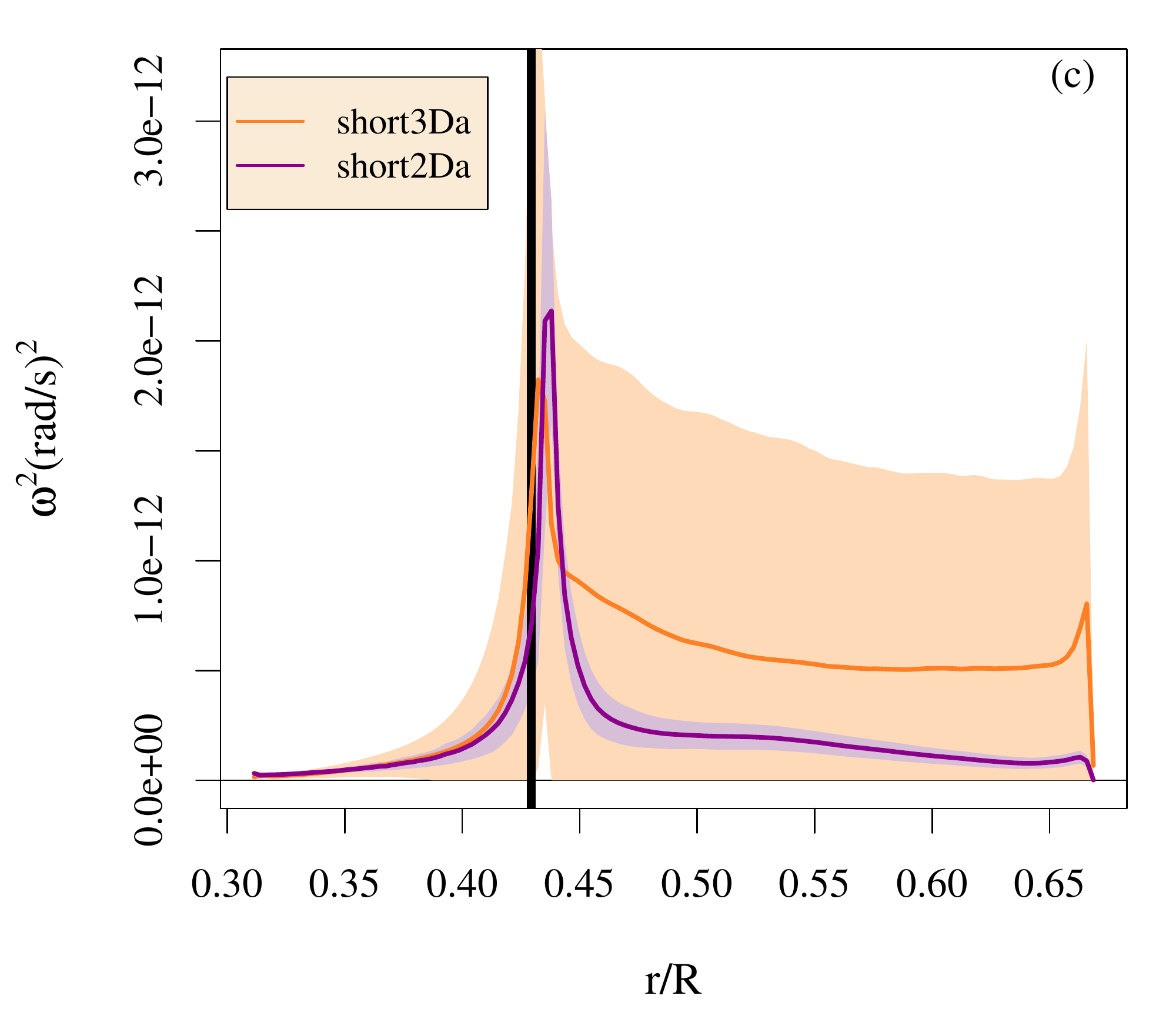}}\resizebox{3.5in}{!}{\includegraphics{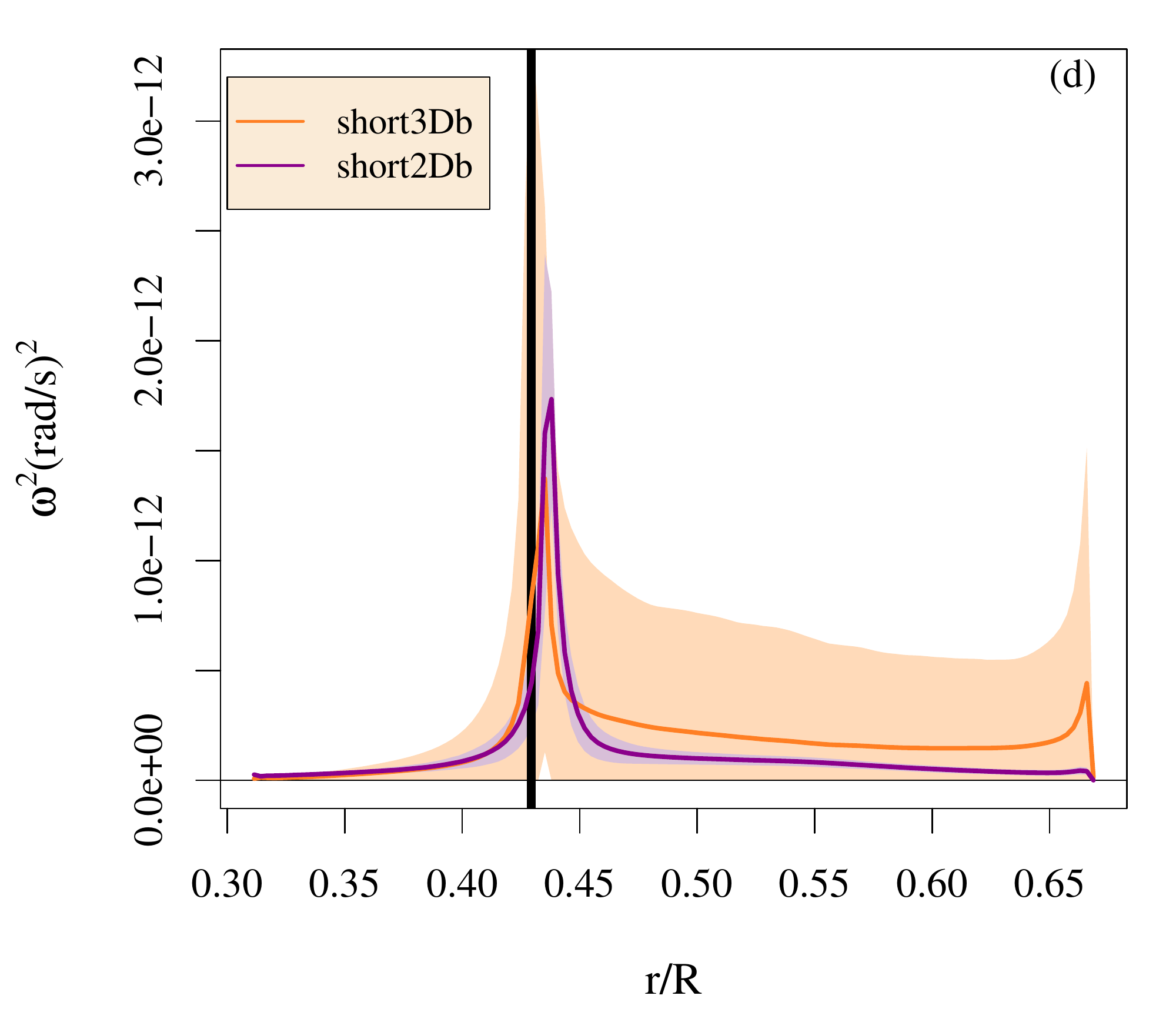}}
\caption{Average radial profile of the local enstrophy (a) in simulations \emph{wide2D} and \emph{wide3D}, (b) in simulations \emph{deep2D} and \emph{deep3D}, (c) in simulations \emph{short2Da} and \emph{short3Da},  (d) in simulations \emph{short2Db} and \emph{short3Db}.  Shaded areas indicate one standard deviation above and below the time-averaged line.  The heavy vertical line indicates the bottom of the convection zone determined by the Schwarzschild criterion.
\label{figvort}}
\end{center}
\end{figure*}

In addition, we examine time-averaged radial profiles of the local enstrophy, obtained from averaging the squared vorticity in the angular directions.  In Fig.~\ref{figvort}, the local enstrophy is compared for our four pairs of simulations.  The local enstrophy is clearly larger in the lower and middle convection zones of our 3D simulations than in our 2D simulations.   This different structuring of the flow may lead to larger local shear and have consequences for mixing properties at the lower boundary of the convection zone.  In the near-surface layers of the wide and deep simulations, the local enstrophy of the 2D and 3D simulations is similar. 
The near-surface layers are not present in the \emph{short-a} and \emph{short-b} simulations.   Nevertheless, in these truncated simulations, the local enstrophy shows a much larger variation in 3D than in 2D.

\section{Results: statistics of convective penetration \label{secpenetration}}

\begin{figure}
\begin{center}
\resizebox{3.5in}{!}{\includegraphics{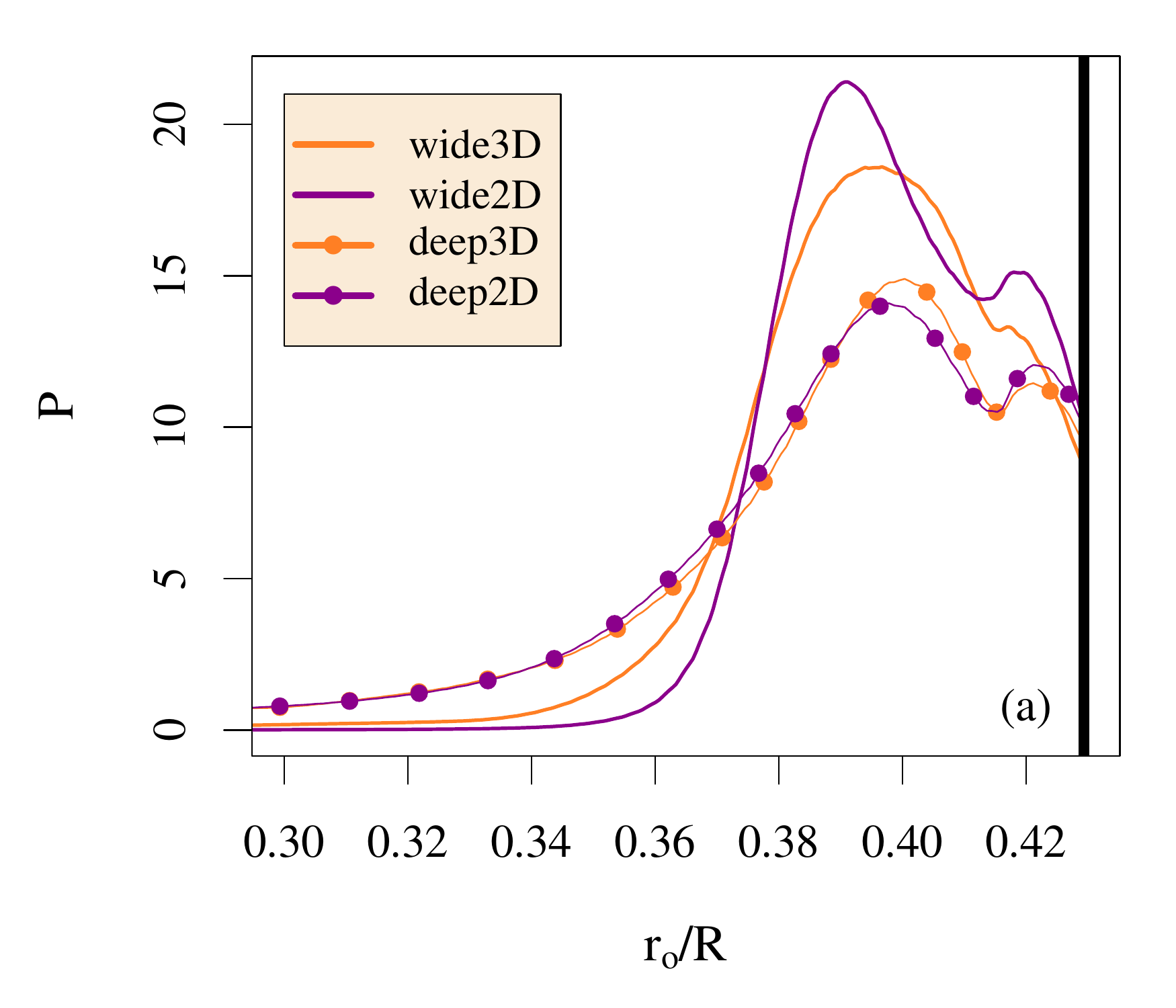}}
\caption{Probability density functions of penetration depth $r_{\mathsf{o}}$ for the \emph{wide} and \emph{deep} simulation pairs determined by the first zero of the vertical kinetic energy flux.  The heavy vertical line indicates the bottom of the convection zone determined by the Schwarzschild criterion.
\label{figkpdf}}
\end{center}
\end{figure}

\begin{figure*}
\begin{center}
\resizebox{5.5in}{!}{\includegraphics{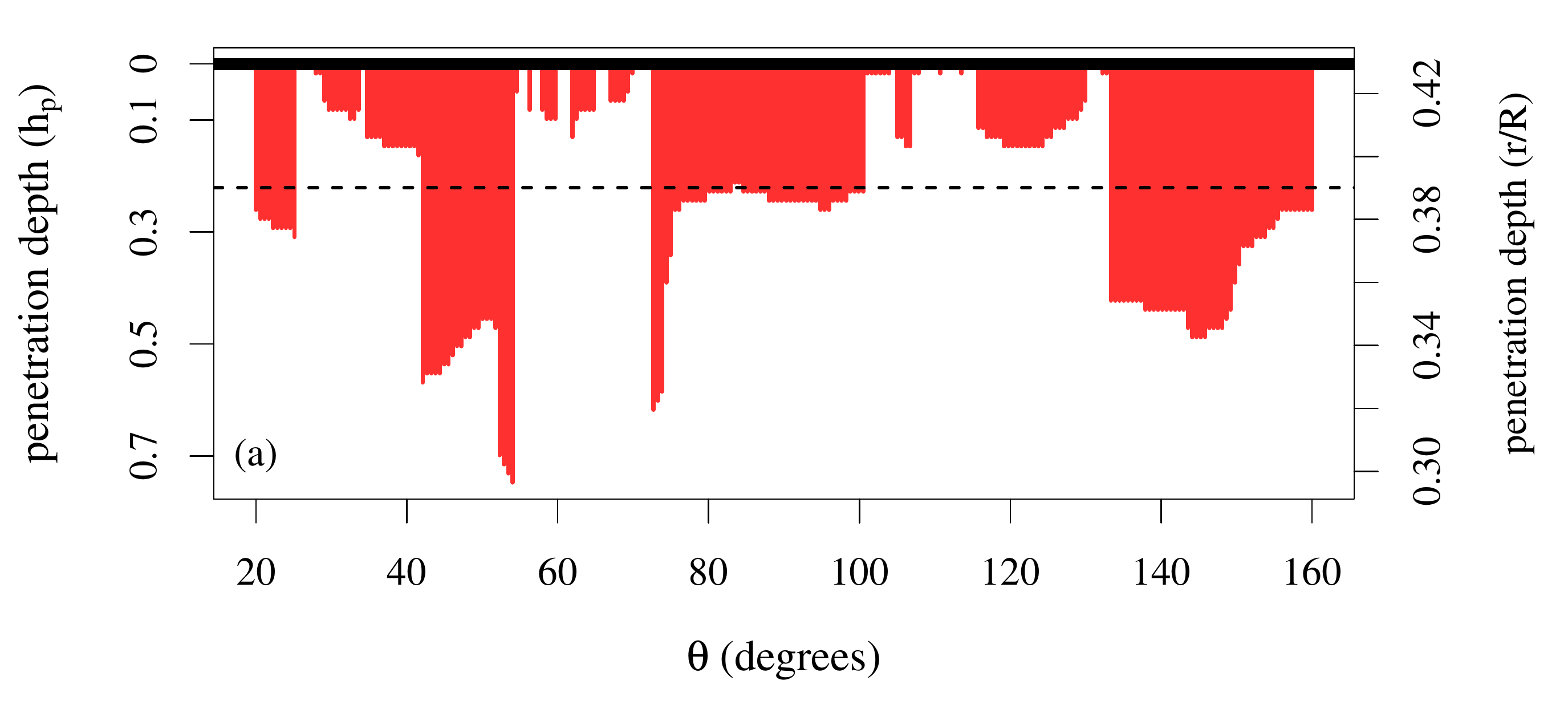}}
\resizebox{5.5in}{!}{\includegraphics{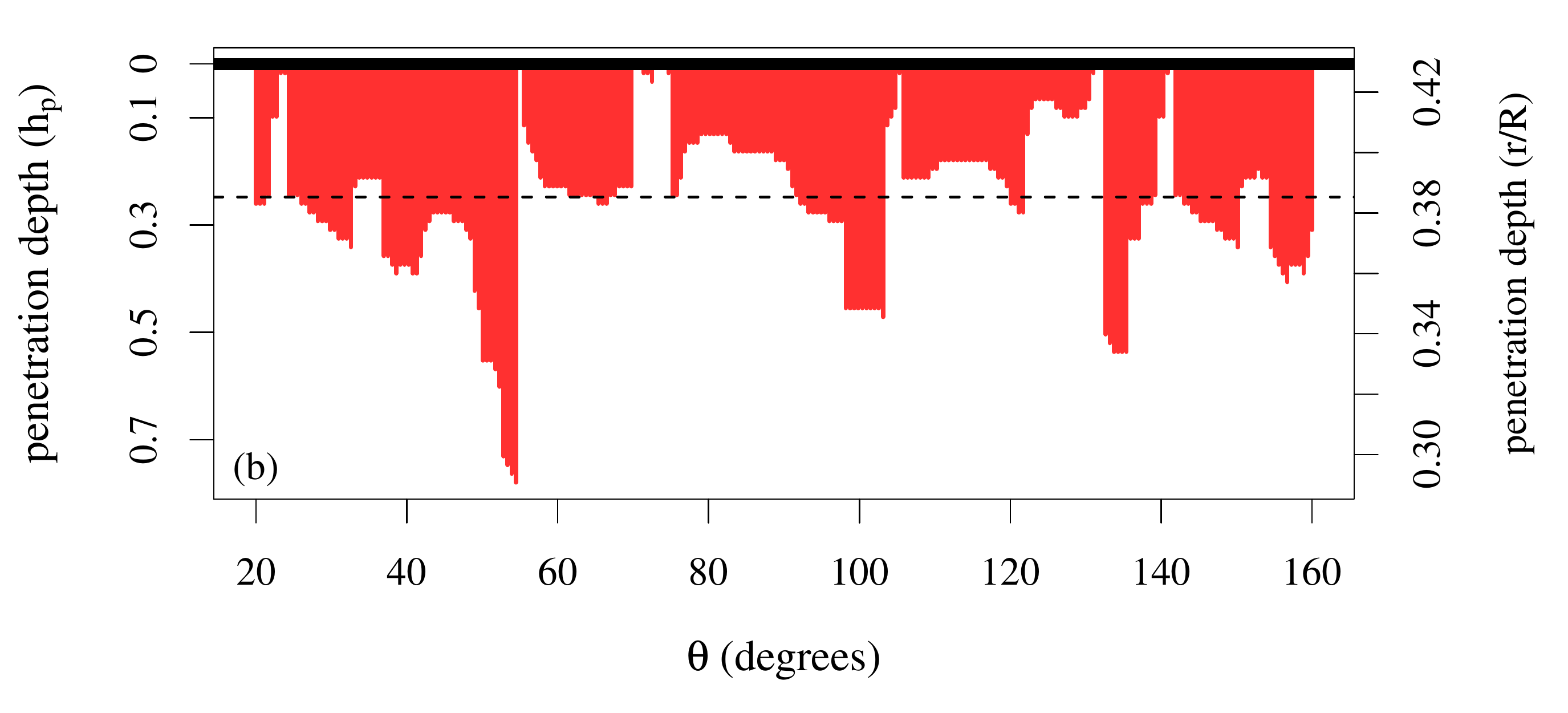}}
\caption{Angular structure of the penetration layer at an arbitrary time in simulation (a) \emph{deep2D} and (b) \emph{deep3D}.  The penetration depth in this illustration is determined by the first zero of the vertical kinetic energy flux.  The boundary between the convection zone and the stable radiative zone, calculated from the Schwarzschild criterion, is indicated by a solid black line.   The vertical axis is in units of the pressure scale height $h_p$ at this boundary.  A dashed black line indicates the average penetration depth at this time.  For the 3D simulation, this 2D representation is of a typical selection in $\phi$.
\label{figred}}
\end{center}
\end{figure*}

Recently we have proposed a new model for determining the width of the penetration layer, based on a statistical analysis of the depth reached by all convective plumes that penetrate below the large convection zone in 2D simulations of the young sun \citep{pratt2017extreme}.  Our model rests on the observation that the statistics of penetration lengths, calculated for each angular grid cell at each time step in our simulation data, produce a strongly non-Gaussian probability distribution, in which the tails of the distribution are of primary physical importance (see Fig.~\ref{figkpdf}).  Because the statistics are non-Gaussian, the use of an average quantity removes critical information about the intermittency of convective penetration; such averaging in both angle and time has been frequently used in early works on this topic. Examples of the structure of the penetration layer are illustrated in Fig.~\ref{figred}.   Moreover, in \citet{pratt2017extreme} we found that when different theoretical measures (vertical kinetic energy flux or vertical heat flux) are used to calculate the point where descending plumes cease, the full probability distributions of penetration depth are similar in form, but the averages can be different.  In that work, we proposed that instead of an average, the \emph{maximum} depth of plume penetration calculated at a single time should be used to define the width of the penetration layer.  Through numerical simulations, we verify that this definition of the penetration layer allows us to accurately pinpoint the depth where penetrating plumes excite waves.   The maximum depth of penetration suggests the application of extreme value theory \citep{castillo2005extreme,charras2013extreme,gomes2015extreme} to determine a form for the diffusion coefficient, enhanced by large-scale convective mixing, to model penetration.  The form of this enhanced diffusion coefficient is:
\begin{eqnarray}\label{newEVTdiffusion}
D(r) = D_0 \mathsf{Pe_{B}}^{1/2} \left( 1- \exp{\left( - \exp{\left(-   \frac{(r_{\mathsf{B}}-r)/R - \mu}{ \lambda} \right)} \right)}  \right)~.
\end{eqnarray}   
Here $\mathsf{Pe_{B}}$ is a characteristic P\'eclet number for the bottom of the convection zone, $r_{\mathsf{B}}$ is the radial position of the convective boundary, and the constants $\lambda$ and $\mu$ are the scale parameter and location parameter of the generalized extreme value distribution (GEVD) which best fits the penetration depth statistics.  Our enhanced diffusion coefficient has been analyzed and applied for stellar evolution calculations \citep{baraffe2017lithium,jorgensen2018addressing,dietrich2018penetrative,augustson2018penetration}.  We refer the reader to \citet{pratt2017extreme} for a complete examination of these statistics and discussion of the development of this model.   The data that inspired this enhanced diffusion coefficient was from two-dimensional simulations of the young sun performed at a range of radial resolutions, and which covered the longer periods of time necessary to produce well-resolved probability density functions to analyze intermittent events.  Here we examine the statistics of convective penetration for 3D simulations, and compare the results to the identically set-up 2D simulations studied in this work.

\begin{figure*}
\begin{center}
\resizebox{3.5in}{!}{\includegraphics{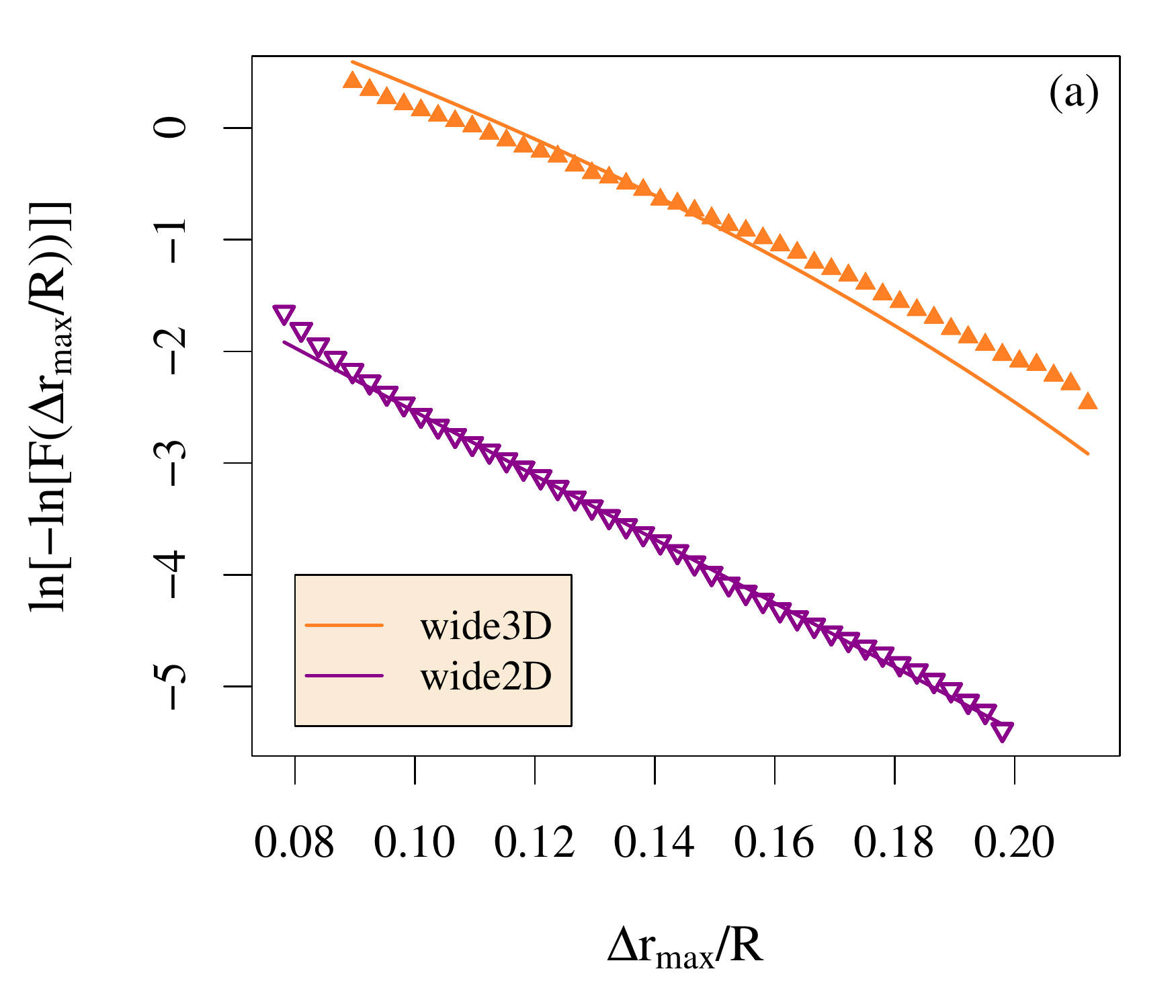}}\resizebox{3.5in}{!}{\includegraphics{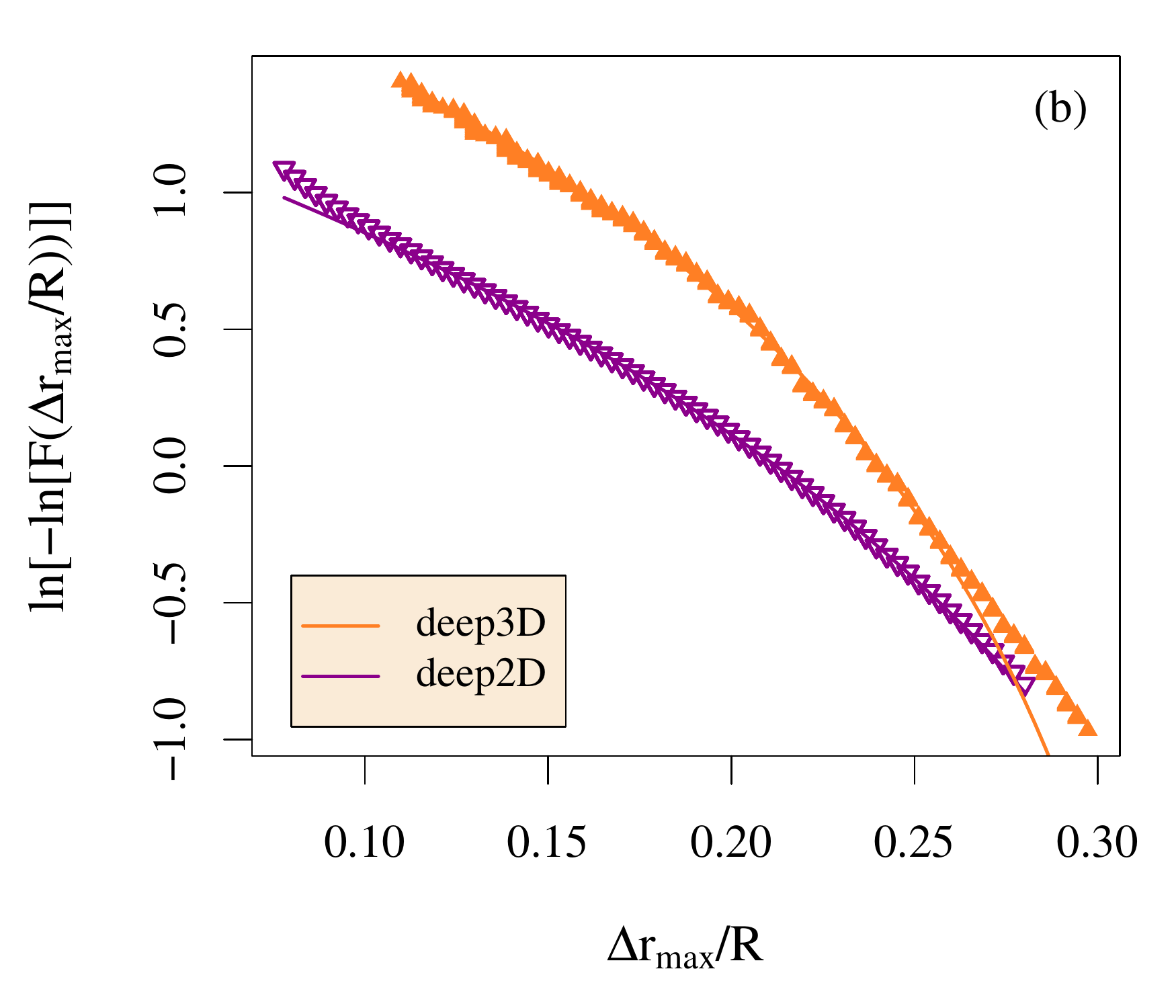}}
\resizebox{3.5in}{!}{\includegraphics{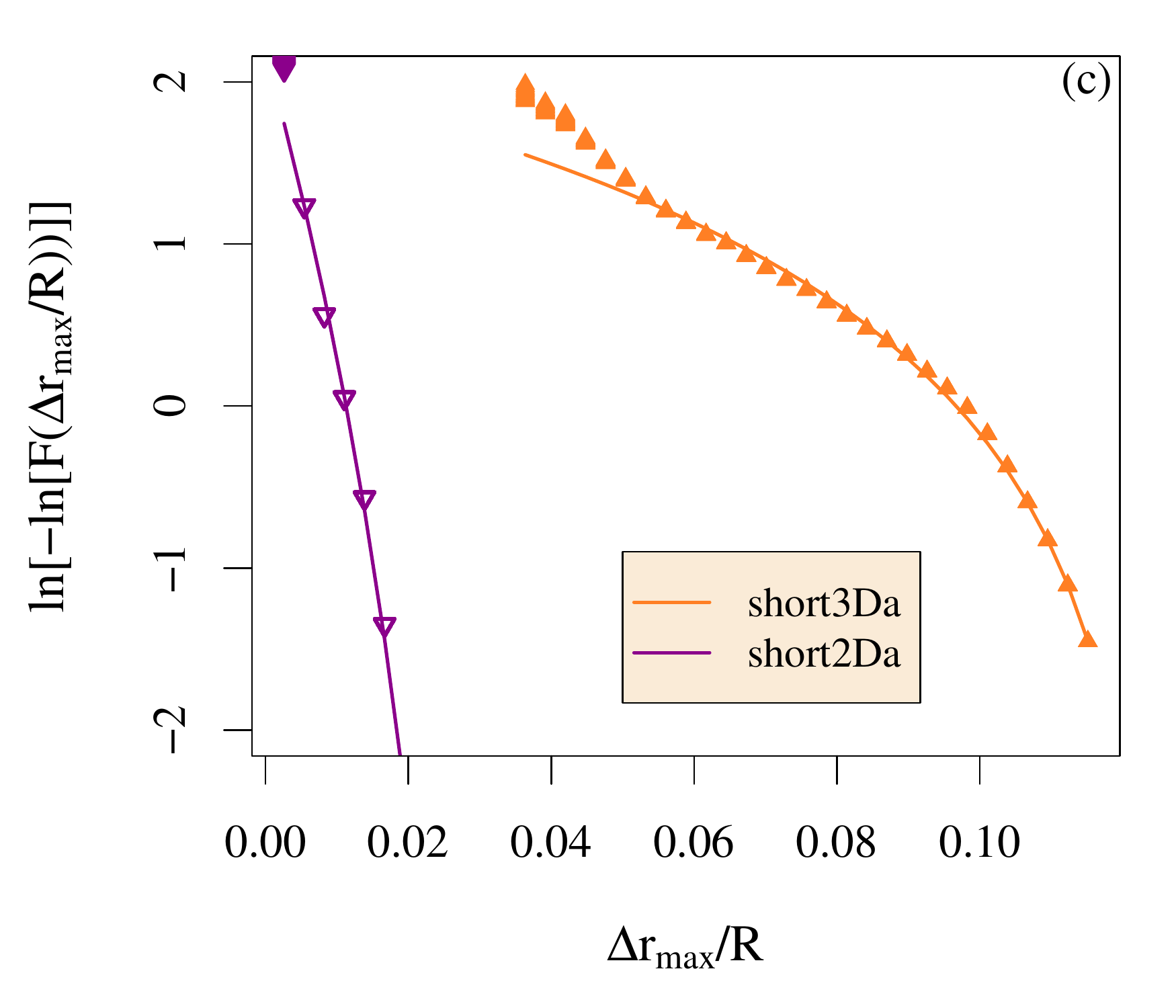}}\resizebox{3.5in}{!}{\includegraphics{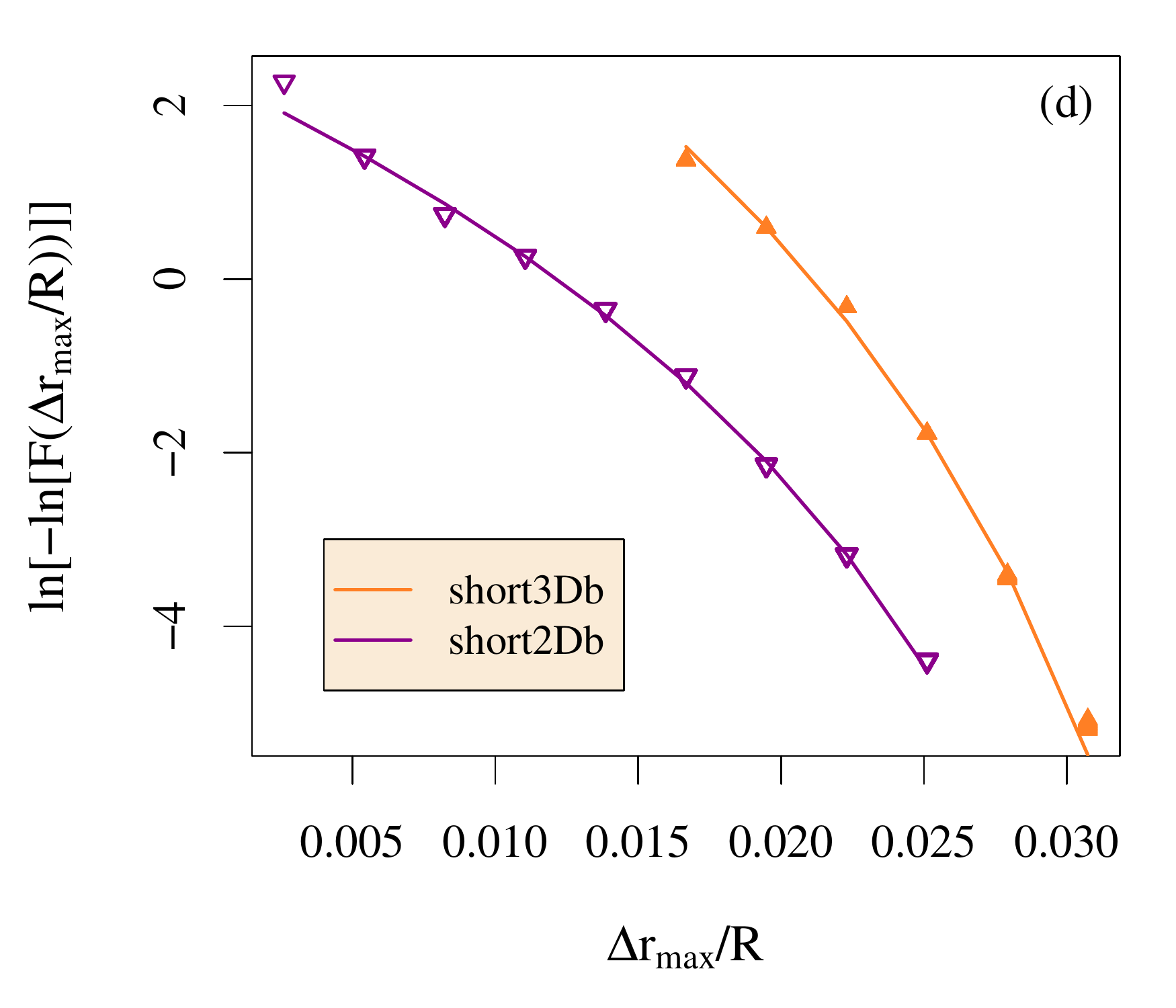}}
\caption{Cumulative distribution function $F$ of the maximal penetration length, $\Delta r_{\mathsf{max}}$ defined from the vertical kinetic energy flux (a) in simulations \emph{wide2D} and \emph{wide3D},  (b) in simulations \emph{deep2D} and \emph{deep3D},  (c) in simulations \emph{short2Da} and \emph{short3Da},  (d) in simulations \emph{short2Db} and \emph{short3Db}.  Triangular points indicate simulation data, while lines show the best for of the GEVD.
\label{figcdffit}}
\end{center}
\end{figure*}

Fig.~\ref{figcdffit} shows the cumulative distribution functions of maximal penetration length for our four pairs of simulations (points), along with the best fit of this data (lines). 
The cumulative distribution function $F$ is equal to the probability $P$ of obtaining a value less than or equal to the argument, i.e. $F(x=A) = P(x \leq A)$, so that all of the information of the probability density function is contained in the cumulative distribution function.  The cumulative distribution function thus describes the accumulative effect of the most vigorous plumes reaching a given depth and characterizes the process of enhanced mixing in this region.
To precisely determine the fit to the GEVD for each of our simulations, we use the package \emph{evd} \citep{revdpackage,penalva2013topics} publicly available for R (The R Project for Statistical Computing).\footnote{ The R Project for Statistical Computing:  https://cran.r-project.org/}  The parameters determined by this fit are summarized in Table~\ref{tablekineticevd}.  In Fig.~\ref{figcdffit}, the natural log of the negative natural log of $F$ is shown; for the simplest kind of GEVD, the Gumbel distribution, this would produce a linear relationship with the maximal penetration length.

Based on the arguments in \citet{pratt2017extreme}, the location parameter $\mu$ of the GEVD can be used as an approximation to a simple overshooting length, i.e. $\ell_{\mathsf{ov}}=\mu$.  The location parameter $\mu$ is larger for each 3D simulation than for each 2D simulation, although that difference varies between simulation pairs.  It is difficult to interpret these differences in a broad sense.  For example, the location parameter for the wide2D simulation is much smaller than the wide3D simulation; however the shape parameters for these simulations indicate that their distributions also have different curvatures, as indicated by the shape parameter so that close comparison of the parameters is difficult.  Because the three-dimensional fit is based on a much shorter period of time than the two-dimensional fit ($\sim 4 \tau_{\mathsf{conv}}$ rather than $> 100 \tau_{\mathsf{conv}}$), it is not clear whether these kinds of difference between 2D and 3D results are statistically significant.  For each pair of simulations, the data fill a short range of values of the maximal penetration length $\Delta r_{\mathsf{max}}/R$, and these ranges largely overlap.  The fitting shows that each simulation is fit best with a negative shape parameter, corresponding to a Weibel distribution in extreme value theory.  In \citet{pratt2017extreme}, we noted that the Weibel distribution of the data appears to converge toward a Gumbel distribution at high radial resolution.  The highest resolution simulations that are feasible in 2D are however not feasible in 3D.

\begin{figure*}
\begin{center}
\resizebox{2.5in}{!}{\includegraphics{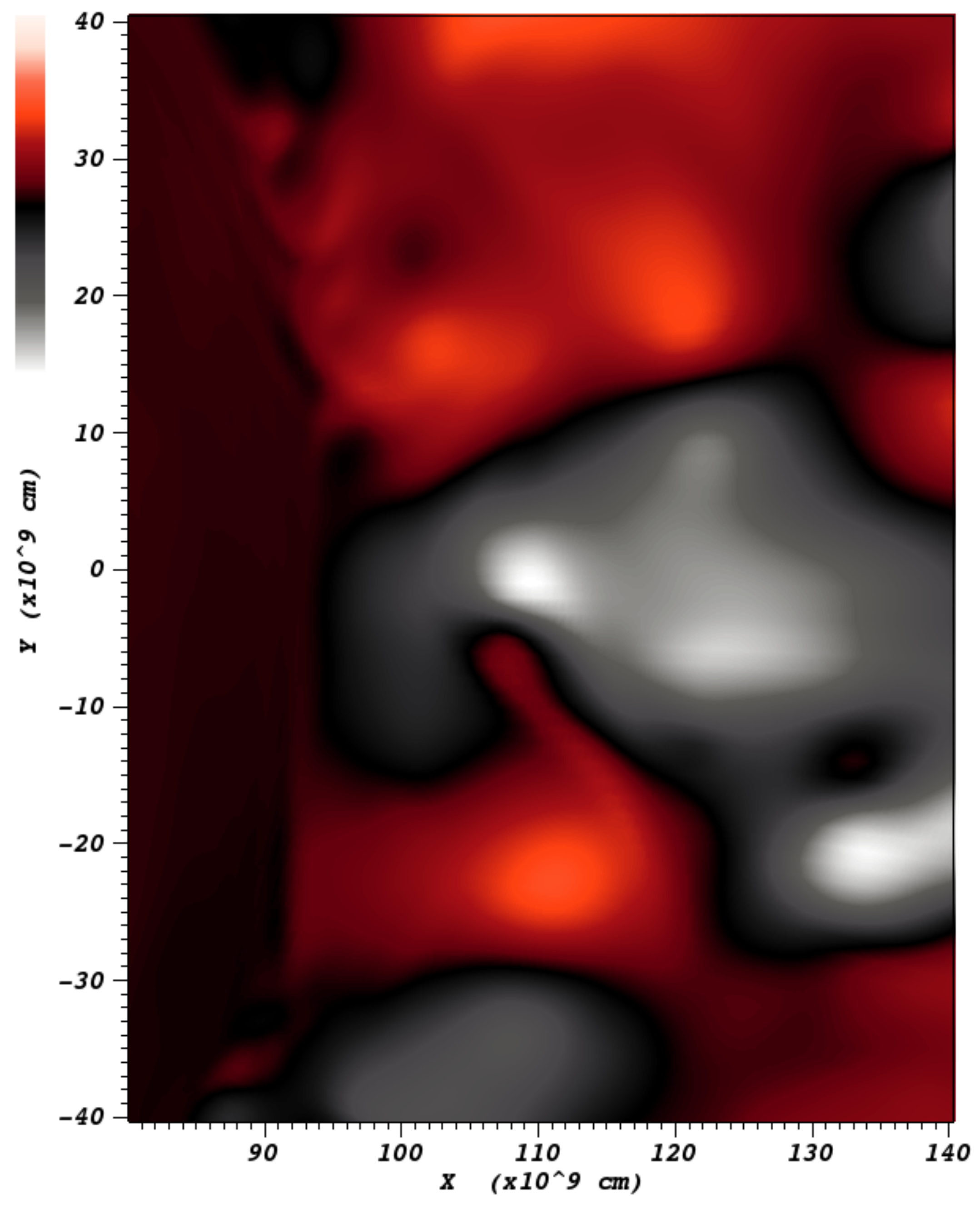}}\resizebox{2.5in}{!}{\includegraphics{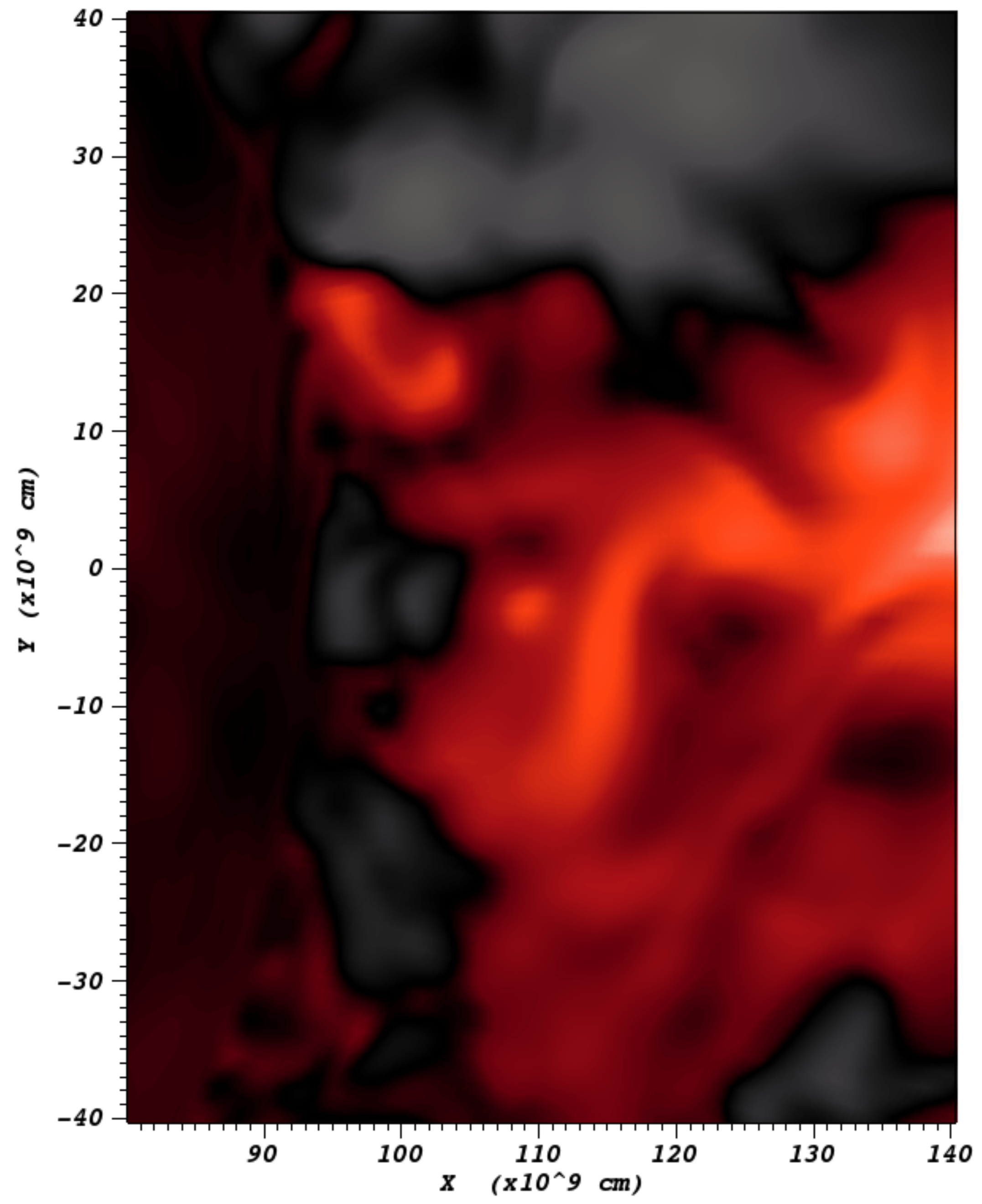}}
\resizebox{2.5in}{!}{\includegraphics{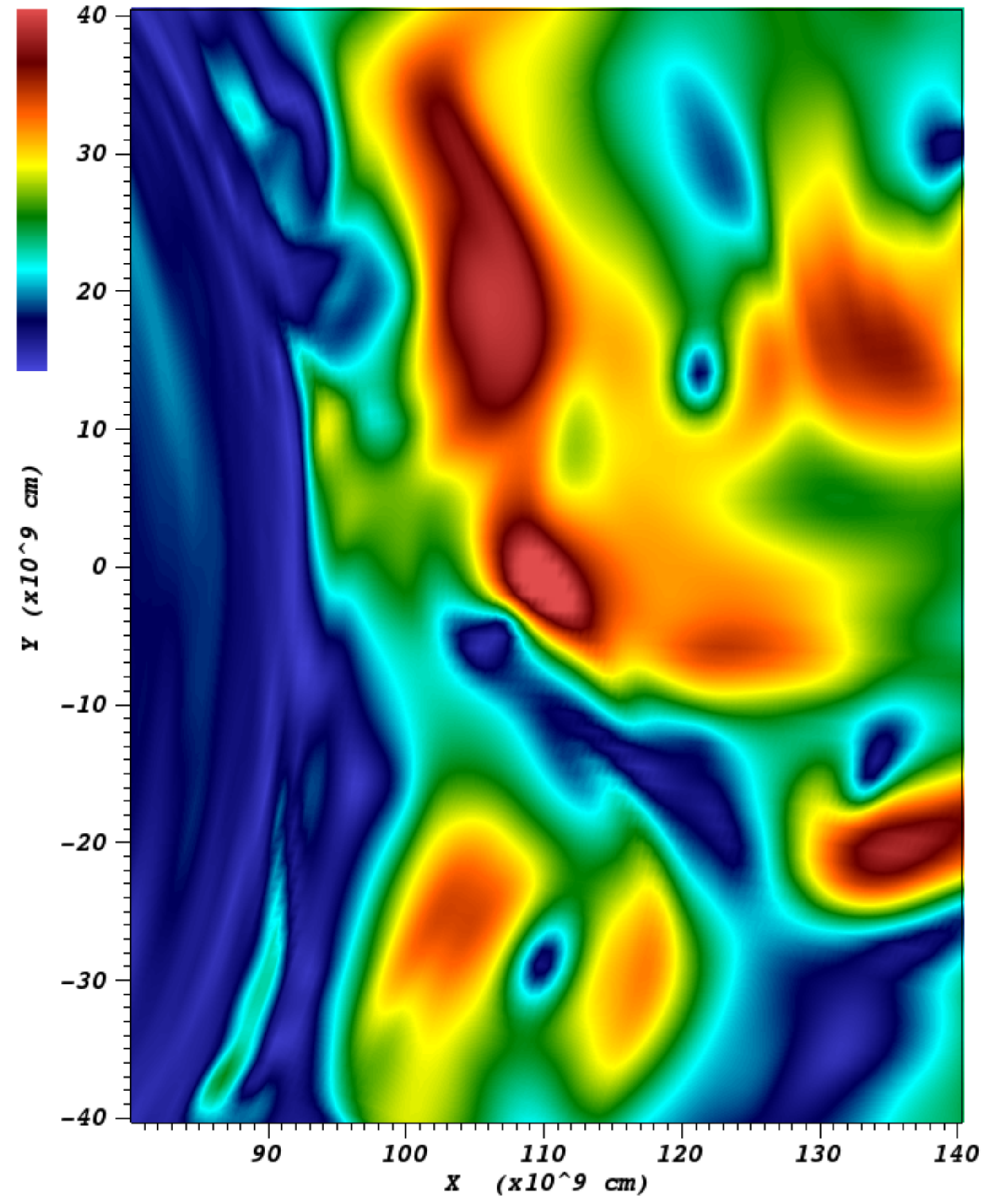}}\resizebox{2.5in}{!}{\includegraphics{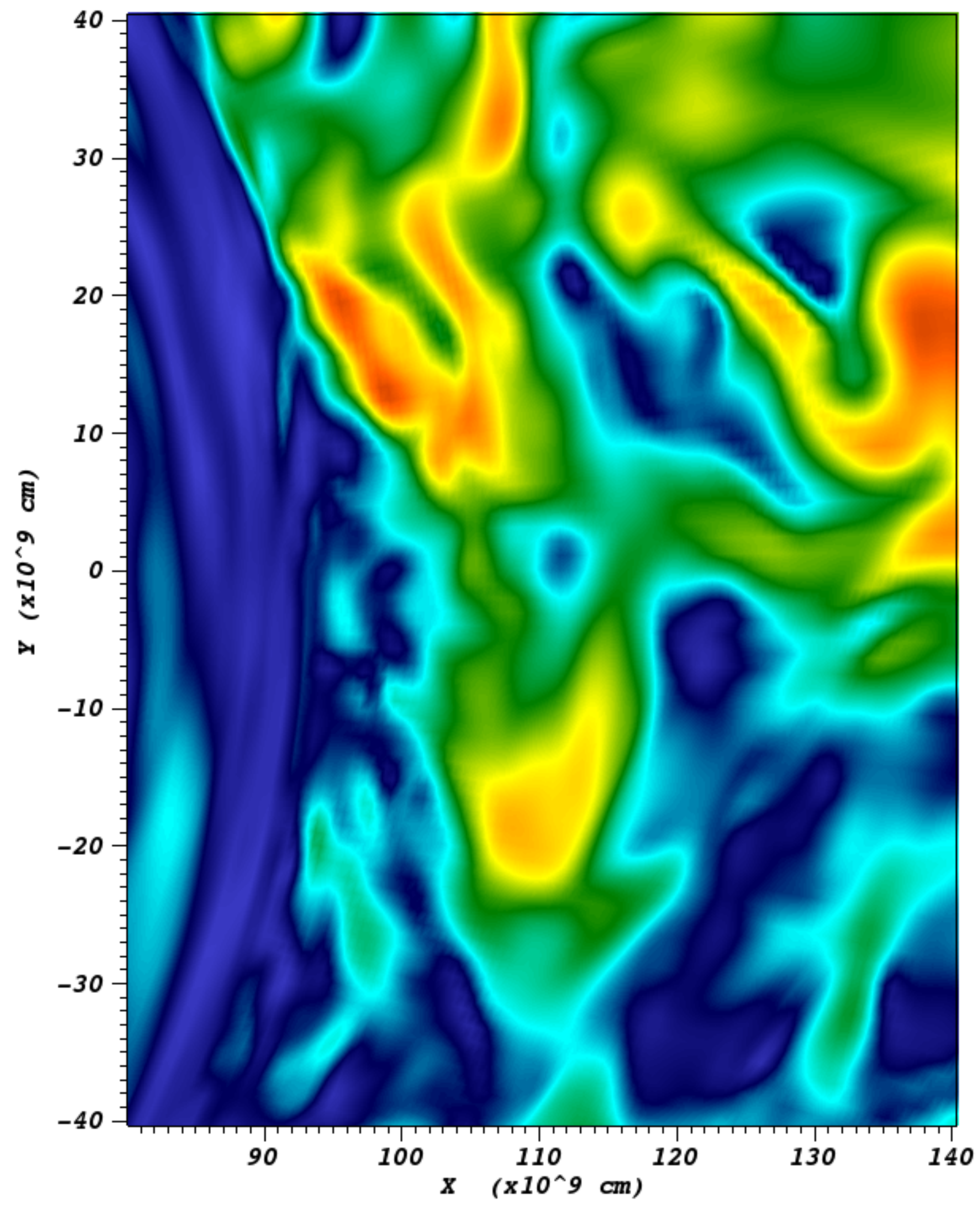}}
\resizebox{2.5in}{!}{\includegraphics{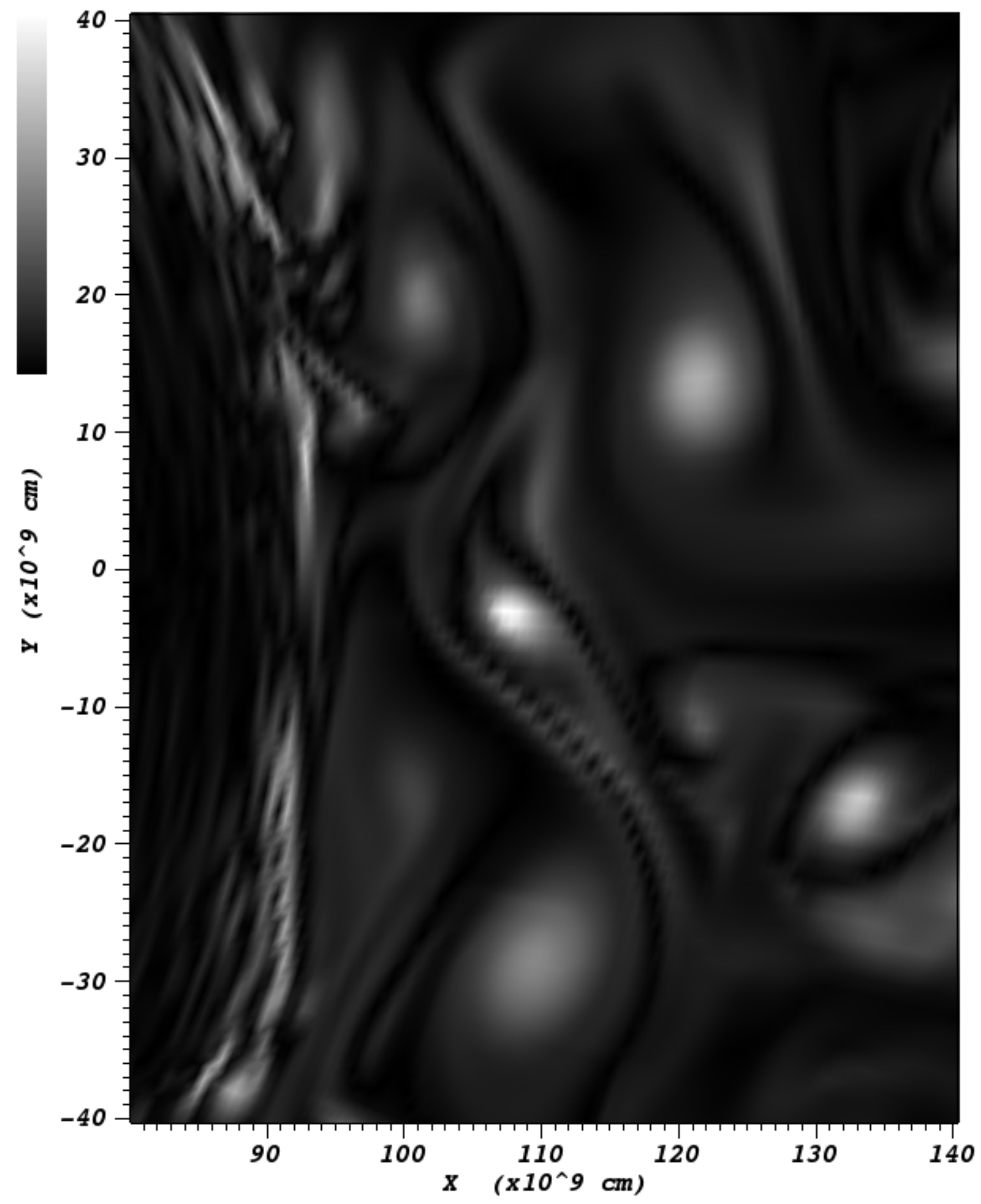}}\resizebox{2.5in}{!}{\includegraphics{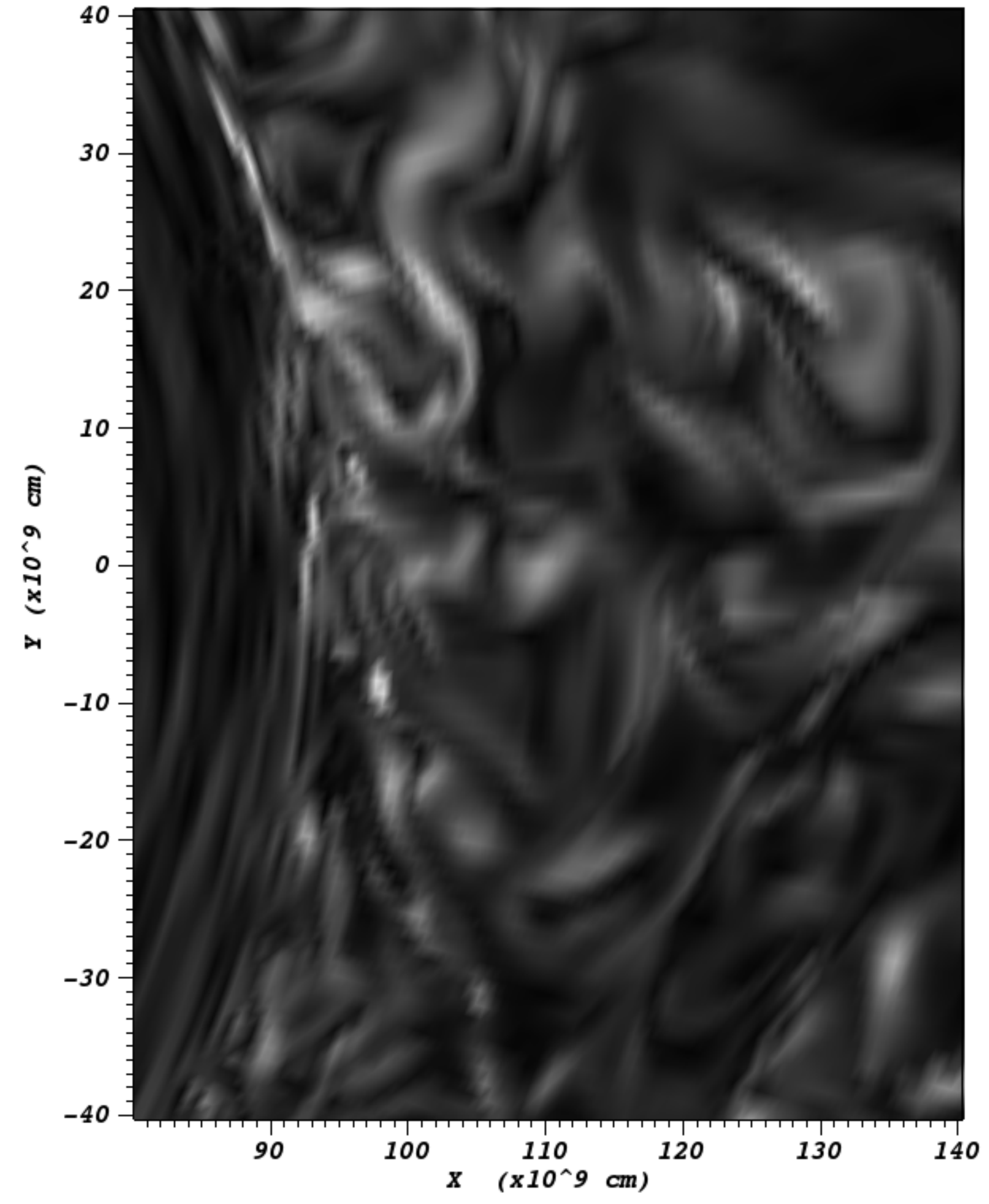}}
\caption{Typical snapshots zoomed in on the same small area of simulation \emph{deep2D} (left), and in a two-dimensional cut of simulation \emph{deep3D} (right).  Horizontal distance spans approximately $80 \cdot 10^9$cm, while vertical distance spans approximately $60 \cdot 10^9$cm. Color scales are identical.   Top: radial velocity, middle: velocity magnitude, bottom: vorticity magnitude.
\label{figzoomviz}}
\end{center}
\end{figure*}

The result that 3D simulations produce a penetration length that is as large as 2D simulations is significant.
Early stellar simulation efforts \citep[e.g.][]{muthsam1995numerical} have reported that 2D simulations have a larger penetration depth than 3D, while in atmospheric convection 2D penetration (in this context termed convective entrainment) has been found to be smaller than in 3D \citep[e.g.]{petch2008differences}.  Although radial velocities $v_r$ in 2D are generally larger than in 3D throughout most of the convection zone, we find that immediately surrounding the convective boundary they have similar average magnitudes.  This presents us with an ambiguity in using the standard analytical model \citep[as discussed by, e.g.][]{schmitt1984overshoot,zahn1991convective,brummell2002penetration,zahn2002convective,brandenburg2016stellar,kapyla2017extended} to predict an ordering of the penetration depth.   This standard model relates the extent of penetration to the exit velocity of the plumes from
the convective region and their filling factor $f$, defined as the fraction of horizontal area occupied by plumes at the edge of the unstable zone.  The formula is $\ell_{\mathsf{ov}} \sim f^{1/2}v_{r,\mathsf{B}}^{3/2}$ where $v_{r,\mathsf{B}}$ is the radial velocity at the convective boundary.\footnote{Total heat flux and thermal diffusivity also play a role in this analytical model \citep[see][for details]{zahn1991convective}.}
Indeed the filling factor has been found to be a more significant predictor for penetration depth than the exit velocity \citep{brummell2002penetration}.
  The filling factor is a quantity that is naturally dependent on the geometry of plumes, and expected to be different in 2D and 3D.  Different shapes of convective structures have been observed in 2D and 3D for plumes in low-to-moderate Prandtl number Rayleigh--B\'enard convection \citep{van2013comparison}.   Another possibility is that interaction between upflows and downflows could be different between 2D and 3D simulations \citep[e.g.][]{rogers2006numerical,rempel2004overshoot}.  
The small scale features and higher vorticity found in our 3D simulations, evident in Figs. \ref{figradvelviz} and \ref{figvortviz}, support both of these ideas. 
   A closer look at the different flow structures in our visualizations (see Fig.~\ref{figzoomviz}) shows strong visual similarities between the size of large-scale structures in radial velocity and velocity magnitudes, and strong differences between the vorticities; these differences are difficult to quantify.
   Several works have suggested that the filling factor and plume geometry should be smaller in 3D than in 2D \citep[e.g. see the discussion in][]{rogers2006numerical}, but no conclusive study of the filling factor and plume shape using the same simulation framework has been performed.    A full quantitative analysis of plume shape, filling factor, and interaction, which is beyond the scope of the present study, would be necessary to develop our result on convective penetration further.


\begin{table*}
\begin{center}
\caption{Parameters for the generalized extreme value distribution of maximal convective penetration length $\Delta r_{\mathsf{max}}$ calculated from the vertical kinetic energy flux.
 \label{tablekineticevd}
 }
\begin{tabular}{lccccccccccccccccccccccccccc}
                      & Location parameter $\mu$ & Scale parameter $\lambda$ & Shape parameter $\kappa$ %
\\ \hline \hline
wide3D                 &  $0.12$   &  $0.042$ &   $-0.17$ 
\\ \hline
wide2D                 &  $0.011$   &  $0.035$ &   $-0.0002$ 
\\ \hline
deep3D                  &  $0.24$ &   $0.059$  &   $-0.62$ 
\\ \hline
deep2D                  &  $0.21$ &   $0.10$  &   $-0.56$ 
\\ \hline
short3Da                &  $0.097$ &  $0.021$ &    $-0.74$
\\ \hline
short2Da                &  $0.011$ &  $0.0042$ &    $-0.18$
\\ \hline
short3Db                & $0.021$  &  $0.0026$  &  $-0.15$ 
\\ \hline
short2Db                & $0.012$  &  $0.0042$  &  $-0.18$ 
\\ \hline \hline
\end{tabular}
\tablefoot{Parameters $\mu$ and $\lambda$ are given in nondimensional units using $R$, the stellar radius, so that the values can be used directly for the diffusion coefficient in eq.~\eqref{newEVTdiffusion}.  The shape parameter $\kappa$ is nondimensional in the GEVD.}
\end{center}
\end{table*}

\section{Summary and Implications \label{secconc}}

This work has focused on extending the results of \citet{prattspherical} and \citet{pratt2017extreme} to three dimensions.  We find that two-dimensional convective velocities are larger on average than three-dimensional velocities, and this ordering holds for both angular and radial velocity components.
The greater amplitudes in velocity produced by two-dimensional simulations are not a clear disadvantage for studies of stellar convection.
Atmospheric convection simulations must reach the correct parameter regimes to predict weather in the atmosphere.
Rayleigh-B\'enard convection simulations are directly comparable with the results of laboratory experiments.
  In contrast to these settings, global simulations of stellar convection currently do not reach Prandtl numbers, Peclet numbers, Rayleigh numbers, and Reynolds numbers accurate to the interior of stars.  Two-dimensional implicit large-eddy simulations explore somewhat less dissipative flows than those produced by three-dimensional simulations.  At the same time two-dimensional simulations can be produced at higher radial resolution than three-dimensional simulations, and cover a longer period of time than three-dimensional simulations.  Thus 2D modeling provides a significant advantage for areas of stellar physics that require higher resolution in a narrow layer of the star, and are dependent on intermittent processes; convective overshooting and convective penetration are topics that benefit from the study of 2D simulations.
  
In their comprehensive review \citet{kupka2017modelling} note that ``2D LES cannot replace 3D LES, if the turbulent
nature of the flow and the detailed geometrical structure of the flow are important or if high quantitative accuracy is needed.''  In contrast to a typical LES method where the effect of the small scales is modeled, in the ILES method turbulence and fingering convection are not modeled, and are not adequately resolved for either 2D or 3D simulations.  ILES are routinely used for global simulations of stellar convection, with the fundamental assumption that the small scales relevant to turbulence do not play an important role; ILES are examined in this work.  Large-scale convective flows clearly dominate the dynamics of our ILES simulations.  This is demonstrated by the radial velocities visualized in Fig.~\ref{figradvelviz}, where the size of the large-scale convective flow structures is similar between our 2D and 3D simulations.  This indicates that the differences between the small-scale dynamics resolved in our 2D and 3D global simulations of the young sun have little effect on the large-scale convective flows.   Thus the first of these conditions, concern about accurately modeling turbulence, cannot be addressed using global simulations of stars studied using an ILES method, regardless of dimensionality.
  Our results support the idea that the second concern of \citet{kupka2017modelling}, regarding the detailed geometrical structure, may have serious implications for convective penetration in stars studied with an ILES method.

 The ratio of averaged $v_{r,\mathsf{RMS}}/v_{\theta,\mathsf{RMS}}$ is generally lower in 2D simulations than in 3D simulations, indicating that in a broad sense, a different geometry is present in the flow on average. 
In comparing average radial profiles of velocity and its radial variations, we find that near the lower convective boundary of the young sun, 2D and 3D simulations produce similar radial velocity amplitudes.  This may contribute to the penetration depths that we calculate, which are as large in 3D as in 2D.
One might predict differences in penetration depth, related to a different shape of convective plumes in 3D. Different shapes are suggested by our visualizations of vorticity, and supported by the generally higher local enstrophy found in our 3D simulations.  This agrees with the study of \citet{van2013comparison} that found the shape and structuring of Rayleigh-B\'enard convection to be different in 2D and 3D simulations at low Prandtl number.

\vspace*{1mm}The difference in geometrical structure may indeed present a disadvantage for two-dimensional ILES of global stellar convection.  
 However, we also find that the effect of using different resolutions, different boundary conditions, or different simulation volumes (related to the extent of the convection zone) has an effect on the velocity and vorticity amplitudes that can be as large or larger than the difference between 2D and 3D results for this type of simulation.  Thus simple differences in simulation set-up result in a different amount of convective penetration: the \emph{deep} simulations and the \emph{short-b} simulations have penetration depths that differ by an order of magnitude (see the horizontal axes of panels (b) and (d) in Fig.~\ref{figcdffit}), although the stratification around the bottom of the convection zone in these simulations is identical.  The sensitivity of convection to the details of the set-up is not surprising, considering the care that is routinely taken for direct numerical simulations of Rayleigh-B\'enard convection, a much more controlled environment.
   This point should be a source of caution for global simulations of stellar convection, since the resolution, aspect ratio, and boundary conditions on the convection zone may create larger or smaller differences between 2D and 3D simulations.  Direct testing of boundary conditions and resolution is likely to be necessary for each numerical model and each physical model of a star to establish the magnitude of differences between 2D and 3D simulations.



\begin{acknowledgements}
The research leading to these results  is partly supported by the ERC grants 320478-TOFU and 787361-COBOM and by the STFC Consolidated
Grant ST/R000395/1.

This work used the DiRAC Complexity system, operated by the University of Leicester IT Services, which forms part of the STFC DiRAC HPC Facility (www.dirac.ac.uk). This equipment is funded by BIS National E-Infrastructure capital grant ST/K000373/1 and STFC DiRAC Operations grant ST/K0003259/1. DiRAC is part of the National E-Infrastructure.

This work also used the University of Exeter local supercomputer ISCA.
\end{acknowledgements}

\bibliographystyle{aa}
\bibpunct{(}{)}{;}{a}{}{,}
\bibliography{music}

\end{document}